\documentclass[12pt]{iopart}
\usepackage{iopams,mathptm,times,graphicx}

\newcounter{theorem}
\renewcommand\thetheorem{\arabic{section}.\arabic{theorem}}

\newenvironment{remark}{\par\medskip\noindent\begingroup{\bf Remark
             \stepcounter{theorem}\thetheorem.}\
             \def\@currentlabel{\thetheorem}}{\endgroup\par\medskip}

\newenvironment{proof}{\par\noindent{\bf Proof.} }{\proofbox\par\medskip}

\def\proofbox{\hfill{\ensuremath\Box}}

\newdimen\LENB \newdimen\LENW \newdimen\THI
\newdimen\LENWH \newdimen\LENTOT \newcount\N
\def\vbrknlnele#1#2#3{
  \LENB=#1pt \LENW=#2pt \THI=#3pt
  \LENWH=\LENW \divide\LENWH by 2
  \LENTOT=\LENB \advance\LENTOT by \LENW
  \vbox to \LENTOT{
    \vbox to \LENWH{}
    \nointerlineskip
    \vbox to \LENB{\hbox to \THI{\vrule width \THI height \LENB}}
    \nointerlineskip
    \vbox to \LENWH{}
  }}

\def\vbrknln#1{
  \N=#1
  \vcenter{
    \vbox{
      \loop\ifnum\N>0
        \vbox to 4pt{\vbrknlnele{2}{2}{0.1}}
        \nointerlineskip
        \advance\N by -1
      \repeat
  }}}

\def\hbrknlnele#1#2#3{
  \LENB=#1pt \LENW=#2pt \THI=#3pt
  \LENTOT=\LENB \advance\LENTOT by \LENW
  \vcenter{
    \vbox to \THI{
      \hbox to \LENTOT{
        \hfil
        \vrule width \LENB height \THI
        \hfil}
  }}}

\makeatletter

\usepackage{ulem}

\def\journal#1&#2,{\begingroup \let\journal=\dummyjournal
               \it #1\unskip~\bf\ignorespaces #2\rm,\endgroup}
\def\dummyjournal{\errmessage{Reference foul up: nested \journal macros}}

\eqnobysec

\def\eqref#1{(\ref{#1})}

\begin{document}
\title[Integrable discretizations and self-adaptive moving mesh method for a CSP equation]
  {Integrable discretizations and self-adaptive moving mesh method for a coupled short pulse equation}
\author{Bao-Feng Feng,
$^1\footnote{Corresponding author; e-mail: feng@utpa.edu}$, Junchao Chen $^2$,
Yong Chen$^{2}$, Ken-ichi Maruno$^{3}$,and Yasuhiro Ohta$^{4}$}

\address{$^1$~Department of Mathematics,
The University of Texas-Rio Grade Valley,
Edinburg, TX 78541
}
\address{$^2$~Shanghai Key Laboratory of Trustworthy Computing, East China
     Normal University, Shanghai 200062, People's Republic of China
}
\address{$^3$~Department of Applied Mathematics, School of Fundamental Science and Engineering,
Waseda University, 3-4-1 Okubo, Shinjuku-ku, Tokyo 169-8555, Japan
}
\address{$^4$~Department of Mathematics,
Kobe University, Rokko, Kobe 657-8501, Japan
}

\date{\today}
\def\submitto#1{\vspace{28pt plus 10pt minus 18pt}
     \noindent{\small\rm To be submitted to : {\it #1}\par}}

\begin{abstract}
In the present paper, integrable semi-discrete and fully discrete analogues of
a coupled short pulse (CSP) equation are constructed.
The key of the construction is the bilinear forms and determinant structure of solutions of the CSP equation.
We also construct $N$-soliton solutions for the semi-discrete and fully discrete analogues of the CSP equations in the form of Casorati determinant. In the continuous limit, we show that the fully discrete CSP equation converges to the semi-discrete CSP equation, then further to the continuous CSP equation.  Moreover, the integrable semi-discretization of the CSP equation is used as a self-adaptive moving mesh method for numerical simulations. The numerical results agree with the analytical results very well.
\par
\par
\kern\bigskipamount\noindent
\today
\end{abstract}

\kern-\bigskipamount


\section{Introduction}
The short pulse (SP) equation
\begin{eqnarray}
 u_{xt}  =  u+ \frac 16 (u^3)_{xx}
\end{eqnarray}
was derived by Sch\"{a}fer and Wayne to describe the propagation of ultra-short optical pulses in nonlinear media \cite{SPE_Org, SPE_CJSW}.
Here, $u=u(x,t)$ represents the magnitude of the electric field, while the subscript $t$ and $x$ denote partial differentiations.
The SP equation represents an alternative approach in contrast with the slowly varying envelope approximation which leads to the nonlinear Sch\"{o}dinger (NLS) equation. As the pulse duration shortens, the NLS equation becomes less accurate, while the SP equation provides an increasingly better approximation to the corresponding solution of the Maxwell equations \cite{SPE_CJSW}. With the rapid progress of ultra-short optical pulse techniques, it is expected that the SP equation and its multi-component generalization will play more and more important roles in applications.

The SP equation has been shown to be completely integrable\cite{Sakovich,Brunelli,Brunelli2}.
The loop-soliton solutions as well as smooth-soliton solutions of the SP solution were found in \cite{Sakovich2,Kuetche,Kuetche2}.
Multi-soliton solutions including multi-loop and multi-breathers ones were given in \cite{Matsuno}.
Periodic solutions to the SP equation were discussed in \cite{Matsuno_SPEreview}.

Similar to the case of the NLS equation \cite{Manakov1974},
it is necessary to consider its two-component or multi-component
generalizations of the SP equation for describing the effects of polarization or anisotropy.
As a matter of fact, several integrable
coupled short pulse have been proposed in the literature \cite%
{PKB_CSPE,Sakovich3,Hoissen_CSPE,Matsuno_CSPE,CSPE_Feng,ZengYao_CSPE}. A complex version of the integrable  coupled short pulse equation in \cite{Hoissen_CSPE,Matsuno_CSPE} is studied in \cite{FengcomplexcomSPE}.
The bi-Hamiltonian structures for the above two-component SP equations were
obtained in \cite{biH_Brunelli}.

Integrable discretizations of soliton equations have received considerable attention
recently \cite{Levi,Grammaticos,Suris,Bobenko}. Integrable semi- and fully discretizations of the SP equation were constructed via Hirota's bilinear method \cite{SPE_discrete1}. The same discretizations were reconstructed from the point view of geometry in \cite{SPE_discrete2}. Most recently, an integrable discretization for a coupled SP equation proposed in \cite{Hoissen_CSPE,Matsuno_CSPE} was constructed  \cite{CSPE_discrete}.


In the present paper, we consider integrable discretizations of another coupled short pulse (CSP) equation
proposed by one of the authors \cite{CSPE_Feng}
\begin{eqnarray}
\label{CSPE1}  u_{xt} & = & u+ \frac 16 (u^3)_{xx}+ \frac 12 v^2 u_{xx},  \\
\label{CSPE2}  v_{xt} & = & v+ \frac 16 (v^3)_{xx}+ \frac 12 u^2 v_{xx}.
\end{eqnarray}

It was shown in \cite{CSPE_Feng} that Eqs. (\ref{CSPE1}) and (\ref{CSPE2}) can be derived from bilinear equations
\begin{eqnarray}
D_sD_y f \cdot f=\frac{1}{2}(f^2-{\bar{f}}^2)\,,\\
D_sD_y \bar{f} \cdot \bar{f}=\frac{1}{2}({\bar{f}}^2-f^2)\,,
\end{eqnarray}

\begin{eqnarray}
D_sD_y g \cdot g=\frac{1}{2}(g^2-{\bar{g}}^2)\,,\\
D_sD_y \bar{g} \cdot \bar{g}=\frac{1}{2}({\bar{g}}^2-g^2)\,,
\end{eqnarray}
through a hodograph transformation
\begin{equation}\label{hodograph}
  x=y -\left(\ln (F\bar{F}) \right)_s\,, \quad t=s\,,
\end{equation}
and dependent variable transformations
\begin{equation}\label{var_trf}
u={\rm i}\left(\ln\frac{\bar{F}}{F}\right)_s\,, \quad
v={\rm i}\left(\ln\frac{\bar{G}}{G}\right)_s\,,
\end{equation}
where $F=fg, G=f\bar{g}$,  $\bar{F}$ and $\bar{G}$
stand for the complex conjugate of $F$ and $G$, respectively.
Meanwhile, $N$-soliton solutions of CSP equation in a parametric form are given,
and the properties of one-soliton, soliton-breather solutions are investigated in detail in \cite{CSPE_Feng}.
The bi-Hamiltonian structure of CSP equation (\ref{CSPE1})-(\ref{CSPE2}) is derived by Brunelli and Sakovich \cite{biH_Brunelli}.

The rest of the present paper is organized as follows.
In section 2, we propose an integrable semi-discrete analogue of the CSP equation via constructing a B\"acklund transform to the bilinear equations of the CSP equation. Meanwhile, $N$-soliton solution is provided in terms of Casorati determinant form. In section 3, starting from two sets of B\"acklund transforms to the bilinear equations of the CSP equation, the fully discrete analogue of the CSP equation is proposed by introducing two auxiliary variables. Moreover, $N$-soliton solution is presented to confirm the integrability. Section 4 is contributed to the self-adaptive moving mesh method by applying the semi-implicit Euler scheme to the semi-discrete CSP.  The paper is concluded by section 5.
Appendix A, B and C present the proofs of Proposition 1, Theorem 1 and Theorem 2, respectively.

\section{Integrable semi-discretization}
We start with two sets of bilinear equations for the semi-discrete two-dimensional Toda-lattice (2DTL) equations with the same discrete parameter $a$
\begin{equation}
\left(\frac{1}{a}D_{x_{-1}}-1\right) \tau_n(k+1)\cdot \tau_n(k)
 +\tau_{n+1}(k+1)\tau_{n-1}(k)= 0\,,
   \label{BK2DTL1}
\end{equation}
\begin{equation}
\left(\frac{1}{a}D_{x_{-1}}-1\right) \tau'_n(k+1)\cdot \tau'_n(k)
 +\tau'_{n+1}(k+1)\tau'_{n-1}(k)= 0\,,
   \label{BK2DTL2}
\end{equation}
which is linked by a B\"acklund transformation \cite{Hirota}
\begin{equation}\label{BT}
(D_{x_{-1}}-1)\tau_n(k)\cdot\tau'_n(k) + \tau_{n+1}(k)\tau'_{n-1}(k)=0.
\end{equation}
\textbf{Proposition 1} The bilinear equations (\ref{BK2DTL1})--(\ref{BT}) admit the following determinant solutions
\begin{equation}
\tau_n(x_{-1},k)
=\left\vert \begin{array}{cccc}
  \phi_n^{(1)}(k) & \phi_{n+1}^{(1)}(k) & \cdots & \phi_{n+N-1}^{(1)}(k) \\
  \phi_n^{(2)}(k) & \phi_{n+1}^{(2)}(k) & \cdots & \phi_{n+N-1}^{(2)}(k) \\
  \cdots & \cdots & \cdots & \cdots \\
  \phi_n^{(N)}(k) & \phi_{n+1}^{(N)}(k) & \cdots & \phi_{n+N-1}^{(N)}(k) \\
  \end{array}\right\vert,
\end{equation}
\begin{equation}
  \tau'_n(x_{-1},k)
  =\left\vert \begin{array}{cccc}
  \psi_n^{(1)}(k) & \psi_{n+1}^{(1)}(k) & \cdots & \psi_{n+N-1}^{(1)}(k) \\
  \psi_n^{(2)}(k) & \psi_{n+1}^{(2)}(k) & \cdots & \psi_{n+N-1}^{(2)}(k) \\
  \cdots & \cdots & \cdots & \cdots \\
  \psi_n^{(N)}(k) & \psi_{n+1}^{(N)}(k) & \cdots & \psi_{n+N-1}^{(N)}(k) \\
  \end{array}\right\vert,
\end{equation}
where
\begin{equation}
\phi_n^{(i)}(k)=p_i^n(1-ap_i)^{-k}e^{\xi_i}+q_i^n(1-aq_i)^{-k} e^{\eta_i}\,,
\end{equation}
\begin{equation}
\psi_n^{(i)}(k)=p_i^n(1-p_i)(1-ap_i)^{-k}e^{\xi_i}+q_i^n (1-q_i) (1-aq_i)^{-k} e^{\eta_i}\,,
\end{equation}
with
$$
\xi_i={p_i}^{-1}x_{-1}+\xi_{i0}\,, \quad \eta_i={q_i}^{-1} x_{-1}+\eta_{i0}\,.
$$
Here $p_i$,  $q_i$, $\xi_{i0}$ and $\eta_{i0}$ are arbitrary parameters which can take either real or complex values.

The proof is presented in Appendix A.

Applying a $2$-reduction condition $q_i=-p_i$, then we could have each of the $\tau$ sequences become a sequence of period $2$, i.e.,
 $\tau_n \Bumpeq  \tau_{n+2}$, $\tau'_n \Bumpeq  \tau'_{n+2}$. Here $\Bumpeq$ means two $\tau$ functions are equivalent up to a constant multiple.  Furthermore, by choosing particular values in phase constants, we can make  $\tau_n$ and $\tau_{n+1}$ complex conjugate to each other. Based on the bilinear equations with $2$-reduction, we construct semi-discrete analogue of the CSP equation by the following theorem:

\textbf{Theorem 1} The following equations constitute an integrable semi-discretization of the coupled short pulse equation (\ref{CSPE1})--(\ref{CSPE2})
\begin{equation} \label{sd_CSP-1}
    \frac{d} {ds} \left( u_{k+1}-u_k \right) =  \frac{1}{2}\delta_k(u_{k+1}+u_k) - \frac{1}{2\delta_k}(u_{k+1}-u_k)(v^2_{k+1}-v^2_k)\,,
\end{equation}

\begin{equation} \label{sd_CSP-2}
    \frac{d} {ds} \left( v_{k+1}-v_k\right) = \frac{1}{2}\delta_k(v_{k+1}+v_k) - \frac{1}{2\delta_k}(v_{k+1}-v_k)(u^2_{k+1}-u^2_k)\,,
\end{equation}
\begin{equation} \label{sd_CSP-3}
 \frac{d \delta_k}{d\,s}=-\frac 12 \left(u^2_{k+1}-u^2_k + v^2_{k+1}-v^2_k \right)\,.
\end{equation}
Furthermore, $N$-soliton solution to the above semi-discrete CSP equation is of the following form
\begin{equation} \label{CSP-sl1}
u_k  = {\rm i}\ln\left(\frac{\bar{f}_k \bar{g}_k}{f_k g_k}\right)_s\,,\ \ v_k  = {\rm i}\ln\left(\frac{\bar{f}_k g_k}{f_k \bar{g}_k}\right)_s\,,
\end{equation}

\begin{equation} \label{CSP-sl2}
x_k=2ka- \left( \ln(f_k \bar{f}_k g_k \bar{g}_k)\right)_s\,,
\end{equation}

\begin{equation} \label{CSP-sl3}
\delta_k =  x_{k+1}-x_k=\frac{a}{2}  \left(\frac{{f}_{k+1}{f}_k}{\bar{f}_{k+1}\bar{f}_k}+\frac{\bar{f}_{k+1}\bar{f}_k}{f_{k+1}f_k} +
   \frac{{g}_{k+1}{g}_k}{\bar{g}_{k+1}\bar{g}_k} + \frac{\bar{g}_{k+1}\bar{g}_k}{g_{k+1}g_k}\right)\,,
\end{equation}
where $f_k$, $g_k$, $\bar{f}_k$ and $\bar{g}_k$ are tau-functions defined by
\begin{equation} \label{CSP-sl4}
 f_k=\tau_0\left(\frac{s}{2},k\right), \ \  \bar{f}_k=\tau_1\left(\frac{s}{2},k\right),\ \ g_k=\tau'_0\left(\frac{s}{2},k\right),\ \ \bar{g}_k=\tau'_1\left(\frac{s}{2},k\right)\,,
\end{equation}
with
\begin{equation} \label{CSP-sl5}
 \tau_n\left(\frac{s}{2},k\right)=\left|\phi^{(i)}_{(n+j-1)}\right|_{1\le i,j\le N}\,, \ \
\tau'_n\left(\frac{s}{2},k\right)=\left|\psi^{(i)}_{(n+j-1)}\right|_{1\le i,j\le N}\,,
\end{equation}
\begin{eqnarray*}
&& \phi_n^{(i)}(k)=p_i^n(1-ap_i)^{-k}{\rm e}^{\frac{1}{2p_i}s+\xi_{i0}}+(-p_i)^n(1+a p_i)^{-k} {\rm e}^{-\frac{1}{2p_i}s+\eta_{i0}}\,,\\
&&\hspace{-1.5cm} \psi_n^{(i)}(k)=p_i^n(1-p_i)(1-ap_i)^{-k}{\rm e}^{\frac{1}{2p_i}s+\xi_{i0}}+(-p_i)^n (1+p_i) (1+ap_i)^{-k} {\rm e}^{-\frac{1}{2p_i}s+\eta_{i0}}\,.
\end{eqnarray*}

The proof is presented in Appendix B.
In the process of the proof, multi-soliton solution expressed in determinant form is obvious.
Next, we show that the semi-discrete CSP equation converges to the CSP equation in the continuous limit.

In the continuous limit $a\to 0$ ($\delta_k\to0$), we have
\begin{equation}
  \frac{u_{k+1}-u_k}{\delta_k}\to \frac{\partial u}{\partial x}\,,
\qquad\frac{u_{k+1}+u_k}{2}\to u\,,
\end{equation}

\begin{equation}
\frac{\partial x}{\partial s}
=\frac{\partial x_0}{\partial s}
+\sum_{j=0}^{k-1}\frac{d\delta_j}{d s}
= -\frac 12\sum_{j=0}^{k-1}(u^2_{j+1}+v^2_{j+1} -u^2_j-v^2_j)
\to -\frac 12 (u^2+v^2)\,.
\end{equation}
Thus

\begin{equation}
 \partial_s=\partial_t+\frac{\partial x}{\partial s}\partial_x
\to\partial_t  -\frac 12 (u^2+v^2) \partial_x\,.
\end{equation}
Consequently, Eq. (\ref{sd_CSP1b}) converges to
$$
\left(\partial_t  -\frac 12 (u^2+v^2) \partial_x \right) u_x = u\left( 1+ u^2_x \right),
$$
which is nothing but the first equation of coupled short pulse equation (\ref{CSPE1}).

It can be shown in the same way that Eq. (\ref{sd_CSP2b}) converges to Eq. (\ref{CSPE2}), the second equation of coupled short pulse equation.

\section{Fully discretizations of the coupled short pulse equation}

To construct a fully discrete analogue of the CSP equation,
we introduce one more discrete variable $l$ which corresponds to
the discrete time variable.

It has been known in \cite{OKMS} that the $\tau$-functions
\begin{equation}
\tau_n(k,l)=\left|\phi^{(i)}_{(n+j-1)}(k,l)\right|_{1\le i,j\le N}\,, \
\tau'_n(k,l)=\left|\psi^{(i)}_{(n+j-1)}(k,l)\right|_{1\le i,j\le N}\,, \
\label{full_taufunction}
\end{equation}
with
\[
\fl
\phi^{(i)}_n(k,l)
=p_i^n(1-ap_i)^{-k}\left(1-b\frac{1}{p_i}\right)^{-l}
e^{\frac{1}{2p_i}s+\xi_{i0}}
+q_i^n(1-aq_i)^{-k}\left(1-b\frac{1}{q_i}\right)^{-l}
e^{\frac{1}{2q_i}s+\eta_{i0}}\,,
\]

\[
\fl
\psi_n^{(i)}(k,l)
=p_i^n(1-p_i)(1-ap_i)^{-k}\left(1-b\frac{1}{p_i}\right)^{-l}
e^{\frac{1}{2p_i}s+\xi_{i0}}
\]
\[+q_i^n (1-q_i) (1-aq_i)^{-k}\left(1-b\frac{1}{q_i}\right)^{-l}
e^{\frac{1}{2q_i}s+\eta_{i0}}\,
\]
satisfy bilinear equations

\begin{equation} \label{Backlund1}
\left(\frac{2}{a}D_s-1\right)\tau_n(k+1,l)\cdot\tau_n(k,l)
+\tau_{n+1}(k+1,l)\tau_{n-1}(k,l)=0\,,
\end{equation}
\begin{equation} \label{Backlund1b}
\left(\frac{2}{a}D_s-1\right)\tau'_n(k+1,l)\cdot\tau'_n(k,l)
+\tau'_{n+1}(k+1,l)\tau'_{n-1}(k,l)=0\,,
\end{equation}
and
\begin{equation} \label{Backlund2}
(2bD_s-1)\tau_{n}(k,l+1)\cdot\tau_{n+1}(k,l)
+\tau_{n}(k,l)\tau_{n+1}(k,l+1)=0\,.
\end{equation}
\begin{equation} \label{Backlund2b}
(2bD_s-1)\tau'_{n}(k,l+1)\cdot\tau'_{n+1}(k,l)
+\tau'_{n}(k,l)\tau'_{n+1}(k,l+1)=0\,.
\end{equation}
Here $n, k, l$ are integers, $a, b$ are real numbers, $p_i, q_i, \xi_{i0}, \eta_{i0}$
are arbitrary complex numbers.

By applying a 2-reduction condition: $q_i=-p_i$, we have
$\tau_{n} \Bumpeq \tau_{n+2}$, $\tau'_{n} \Bumpeq \tau'_{n+2}$. We can further have
$$
f_{k,l}=\tau_0(k,l)\,,\quad \bar{f}_{k,l}=\tau_{1}(k,l)\,, \quad
g_{k,l}=\tau'_0(k,l)\,,\quad \bar{g}_{k,l}=\tau'_{1}(k,l)\,,
$$
by adjusting phases in $\phi_i^{(n)}(k,l)$ and $\psi_i^{(n)}(k,l)$. Here $\bar{f}$ and $\bar{g}$ represent complex conjugate functions of $f$ and $g$, respectively. A fully discrete CSP equation can be constructed as follows:

\textbf{Theorem 2} The fully discrete analogue of the CSP equation (\ref{CSPE1})--(\ref{CSPE2}) is of the form
\begin{eqnarray}
\label{fcsp-19} \nonumber && \frac{1}{b}( u_{k+1,l+1} {-} u_{k+1,l} {-} u_{k,l+1} {+} u_{k,l} {+} v_{k+1,l+1} {-} v_{k+1,l} {-} v_{k,l+1} {+} v_{k,l} )\\
\nonumber &=& (y_{k+1,l} {-} y_{k,l+1})( u_{k+1,l+1} {+} u_{k,l} {+} v_{k+1,l+1} {+} v_{k,l} ) \\
&&+ (y_{k+1,l+1} {-} y_{k,l})(u_{k,l+1} {+} u_{k+1,l} {+} v_{k,l+1} {+} v_{k+1,l})\,,\
\end{eqnarray}
\begin{eqnarray}
\label{fcsp-20} \nonumber && \frac{1}{b}( u_{k+1,l+1} {-} u_{k+1,l} {-} u_{k,l+1} {+} u_{k,l} {-} v_{k+1,l+1} {+} v_{k+1,l} {+} v_{k,l+1} {-} v_{k,l})\\
\nonumber & = & (z_{k+1,l} {-} z_{k,l+1})( u_{k+1,l+1} {+} u_{k,l} {-} v_{k+1,l+1} {-} v_{k,l} ) \\
&&+ (z_{k+1,l+1} {-} z_{k,l})(u_{k,l+1} {+} u_{k+1,l} {-} v_{k,l+1} {-} v_{k+1,l})\,,\
\end{eqnarray}
\begin{eqnarray}
\label{fcsp-21} \nonumber && (y_{k+1,l+1}-y_{k+1,l}-y_{k,l+1}+y_{k,l})\Big(\frac{1}{b} + y_{k,l+1}-y_{k+1,l}\Big)\\
\nonumber & =& {-} \frac{1}{4}(u_{k,l+1} {+} u_{k+1,l} {+} v_{k,l+1} {+} v_{k+1,l}) \\
&&\times(u_{k+1,l+1} {+} u_{k+1,l} {-} u_{k,l+1} {-} u_{k,l} {+} v_{k+1,l+1} {+} v_{k+1,l} {-} v_{k,l+1} {-} v_{k,l})\,,
\end{eqnarray}
\begin{eqnarray}
\label{fcsp-22} \nonumber && (z_{k+1,l+1}-z_{k+1,l}-z_{k,l+1}+z_{k,l})\Big(\frac{1}{b} + z_{k,l+1}-z_{k+1,l}\Big)\\
\nonumber & = & -\frac{1}{4}(u_{k,l+1} {+} u_{k+1,l} {-} v_{k,l+1} {-} v_{k+1,l})  \\
&& \times(u_{k+1,l+1} {+} u_{k+1,l} {-} u_{k,l+1} {-} u_{k,l} {-} v_{k+1,l+1} {-} v_{k+1,l} {+} v_{k,l+1} {+} v_{k,l})\,,
\end{eqnarray}
where
\begin{equation}
\label{full_vartrf}
u_{k,l}= {\rm i}\ln\left(\frac{\bar{f}_{k,l} \bar{g}_{k,l}}{f_{k,l} g_{k,l}}\right)_s\,, \quad
v_{k,l}= {\rm i}\ln\left(\frac{\bar{f}_{k,l} g_{k,l}}{f_{k,l} \bar{g}_{k,l}}\right)_s\,,
\end{equation}

\begin{equation}
\label{full_hodograph1}
y_{k,l}=ka-(\ln(f_{k,l} \bar{f}_{k,l})  )_s,\ \ z_{k,l}=ka-(\ln (g_{k,l} \bar{g}_{k,l}) )_s\,,
\end{equation}
and
\begin{equation}
\label{full_hodograph2}
x_{k,l}=y_{k,l}+z_{k,l} =2ka- \left(\ln(f_{k,l} \bar{f}_{k,l} g_{k,l} \bar{g}_{k,l})\right)_s \,.
\end{equation}

The proof is presented in Appendix C.

Finally we show that Eqs. (\ref{fcsp-19})--(\ref{fcsp-22}) converge to the semi-discrete CSP equations
(\ref{sd_CSP-1})--(\ref{sd_CSP-3}) by taking a continuous limit in time ($b\to 0$). Under this limit, Eqs. (\ref{fcsp-19})--(\ref{fcsp-22}) become
\begin{eqnarray}
\label{fcsp-23} && \frac{d}{ds}(u_{k+1}-u_k)+\frac{d}{ds}(v_{k+1}-v_k)=(y_{k+1}-y_k)(u_{k+1}+u_k+v_{k+1}+v_k),\\
\label{fcsp-24} && \frac{d}{ds}(u_{k+1}-u_k)-\frac{d}{ds}(v_{k+1}-v_k)=(z_{k+1}-z_k)(u_{k+1}+u_k-v_{k+1}-v_k), \\
\label{fcsp-25} && \frac{d}{ds}(y_{k+1}-y_k)=-\frac{1}{4}(u_{k+1}+u_k+v_{k+1}+v_k)(u_{k+1}-u_k+v_{k+1}-v_k),\\
\label{fcsp-26} && \frac{d}{ds}(z_{k+1}-z_k)=-\frac{1}{4}(u_{k+1}+u_k-v_{k+1}-v_k)(u_{k+1}-u_k-v_{k+1}+v_k),
\end{eqnarray}
where $\frac{F_{l+1}-F_l}{2b}\rightarrow\partial_sF$($b\rightarrow 0$) is used. Obviously, we have
\begin{equation}\label{fcsp-27}
\frac{d}{ds}(u_{k+1}-u_k)=\frac{1}{2}\delta_k(u_{k+1}+u_k)-\frac{1}{2}(v_{k+1}+v_k)(z_{k+1}-z_{k}-y_{k+1}+y_k)\,,\ \
\end{equation}

\begin{equation}\label{fcsp-28}
\frac{d}{ds}(v_{k+1}-v_k)=\frac{1}{2}\delta_k(v_{k+1}+v_k)-\frac{1}{2}(u_{k+1}+u_k)(z_{k+1}-z_{k}-y_{k+1}+y_k)\,,\ \
\end{equation}
from (\ref{fcsp-23})--(\ref{fcsp-24}), and
\begin{equation}\label{fcsp-29}
\frac{d}{ds} (x_{k+1}-x_k)=-\frac{1}{2}(u^2_{k+1}-u^2_k+v^2_{k+1}-v^2_k),
\end{equation}
by adding (\ref{fcsp-25}) and (\ref{fcsp-26}). Eq. (\ref{fcsp-29}) coincides with Eq. (\ref{sd_CSP-3})

Finally, in view of the relations (\ref{fcsp-01})--(\ref{fcsp-06}), we have
\begin{eqnarray}
&&(z_{k+1}-z_{k}-y_{k+1}+y_k)\delta_k \nonumber\\
&=&(z_{k+1}-z_{k}-y_{k+1}+y_k)(z_{k+1}-z_{k}+y_{k+1}-y_k) \nonumber\\
&=&-\frac{a^2}{4}\Big[\Big({ \frac {{\bar{f}_{k + 1 }}\,{\bar{f}
_{k}}}{{f_{k + 1}}\,{f_{k}}}}    + { \frac {{f_{k + 1}}\,{f_{k}}}{{\bar{f}_{k + 1}}\,{\bar{f}_{k}}}}\Big)^2   -  \Big({  \frac {{\bar{g}_{k + 1}}\,{\bar{g}_{k}}}{{g_{k
 + 1}}\,{g_{k}}}} + { \frac {{g_{k + 1}}\,{g_{k}}
}{{\bar{g}_{k + 1}}\,{\bar{g}_{k}}}}\Big)^2  \Big] \nonumber\\
&=&(u_{k+1}-u_k)(v_{k+1}-v_k).
\label{fcsp-30}
\end{eqnarray}
 A substitution of (\ref{fcsp-30}) into (\ref{fcsp-27})--(\ref{fcsp-28}) yields  (\ref{sd_CSP-1})--(\ref{sd_CSP-2}).

From the construction of the fully discrete analogue of the CSP equation, the multi-soliton solution can be expressed in the following determinant form
\begin{eqnarray}
\hspace{-1cm} &&u_{k,l}   = {\rm i}\ln\left(\frac{\bar{f}_{k,l} \bar{g}_{k,l}}{f_{k,l} g_{k,l}}\right)_s=\frac {\rm i}2 \left( \frac{\bar{f'}_{k,l} }{\bar{f}_{k,l} } + \frac{ \bar{g'}_{k,l}}{\bar{g}_{k,l}} - \frac{{f'}_{k,l} }{f_{k,l} } -\frac{{g'}_{k,l} }{g_{k,l} }  \right)  \,,\\
\hspace{-1cm} &&v_{k,l}   = {\rm i}\ln\left(\frac{\bar{f}_{k,l} g_{k,l}}{f_{k,l} \bar{g}_{k,l}}\right)_s=\frac {\rm i}2  \left( \frac{\bar{f'}_{k,l} }{\bar{f}_{k,l} } +\frac{{g'}_{k,l} }{g_{k,l}}  - \frac{{f'}_{k,l} }{f_{k,l} } - \frac{ \bar{g'}_{k,l}}{\bar{g}_{k,l}}   \right)  \,,\\
\hspace{-1cm} && y_{k,l}=ka-(\ln(f_{k,l} \bar{f}_{k,l})  )_s=ka-\frac 12 \left( \frac{\bar{f'}_{k,l} }{\bar{f}_{k,l} } +\frac{{f'}_{k,l} }{f_{k,l} }   \right),\\
\hspace{-1cm} && z_{k,l}=ka-(\ln (g_{k,l} \bar{g}_{k,l}) )_s=ka-\frac 12 \left( \frac{ \bar{g'}_{k,l}}{\bar{g}_{k,l}}  +\frac{{g'}_{k,l} }{g_{k,l} }  \right),
\end{eqnarray}
thus
\begin{equation}
\hspace{-1.2cm}x_{k,l}=2ka- \left(\ln(f_{k,l} \bar{f}_{k,l} g_{k,l} \bar{g}_{k,l})\right)_s = 2ka- \frac 12 \left( \frac{\bar{f'}_{k,l} }{\bar{f}_{k,l} } + \frac{ \bar{g'}_{k,l}}{\bar{g}_{k,l}} +\frac{{f'}_{k,l} }{f_{k,l} } +\frac{{g'}_{k,l} }{g_{k,l} }  \right)  \,.
\end{equation}
Here
\begin{equation}
f_{k,l}=\tau_0(k,l)\,,\ \  \bar{f}_{k,l}=\tau_{1}(k,l)\,,\ \
g_{k,l}=\tau'_0(k,l)\,,\ \  \bar{g}_{k,l}=\tau'_{1}(k,l)\,,
\end{equation}
\begin{equation}
f'_{k,l}=\rho_0(k,l)\,,\ \ \bar{f'}_{k,l}=\rho_{1}(k,l)\,,\ \
g'_{k,l}=\rho'_0(k,l)\,,\ \ \bar{g'}_{k,l}=\rho'_{1}(k,l)\,,
\end{equation}
with $\tau_n(k,l)$ and $\tau'_n(k,l)$ defined the same as (\ref{full_taufunction}), $\rho_n(k,l)$ and  $\rho'_n(k,l)$ defined as
\begin{equation}
\rho_n(k,l)=\left|\phi^{(i)}_{(n+j-2)}(k,l)\right|_{1\le i,j\le N}\,, \
\rho'_n(k,l)=\left|\psi^{(i)}_{(n+j-3)}(k,l)\right|_{1\le i,j\le N}\,,
\end{equation}
under the 2-reduction condition $q_i=-p_i$ ($i=1, \cdots, N$).

\begin{remark}
Two intermediate variables $y_k$ and $z_k$ are introduced in constructing the fully discrete CSP equation. This often happens when we construct the fully discretizations of  the coupled system such as the coupled modified KdV equation \cite{dcmKdVHirota}.
\end{remark}
\section{Integrable self-adaptive moving mesh method}
In this section, we propose a self-adaptive moving mesh method for the CSP equation (\ref{CSPE1})--(\ref{CSPE2}) and demonstrate the advantage of this integrable scheme by performing several numerical experiments.
\subsection{Numerical scheme}
One of the self-adaptive moving mesh methods for the coupled short pulse equation can be constructed by applying a semi-implicit Euler scheme to its integrable semi-discrete CSP equations (\ref{sd_CSP-1})-- (\ref{sd_CSP-3}). The resulting numerical scheme reads

\begin{equation} \label{mvmesh_CSP-1}
p^{n+1}_k  =  p^{n}_k +\frac {\Delta t}{2} \delta^{n}_k (u^{n}_{k+1}+u^{n}_k)  - \frac{\Delta t}{2\delta^{n}_k} p^{n}_k q^{n}_k (v^{n}_{k+1}+v^{n}_k)\,,
\end{equation}

\begin{equation} \label{mvmesh_CSP-2}
   q^{n+1}_k  =  q^{n}_k +\frac {\Delta t}{2} \delta^{n}_k (v^{n}_{k+1}+v^{n}_k)  - \frac{\Delta t}{2\delta^{n}_k} p^{n}_k q^{n}_k (u^{n}_{k+1}+u^{n}_k)\,,
\end{equation}
\begin{equation} \label{mvmesh_CSP-3}
 \delta^{n+1}_k= -\frac {\Delta t}{2} \left((u^{n+1}_{k+1})^2+(v^{n+1}_{k+1})^2- (u^{n+1}_k)^2 -(v^{n+1}_k)^2 \right)\,.
\end{equation}
Here $p_k=u_{k+1}-u_k$, $q_k=v_{k+1}-v_k$, $\delta_k=x_{k+1}-x_k$. The superscript $n$ represents the numerical value at $t=n \Delta t$. The periodic boundary condition is applied. For convenience, we reserve the time $t \to -t$ so that the left-moving wave becomes right-moving one. In what follows, we report the numerical results for one- and two- soliton solutions.

\subsection{Numerical experiments}
For the sake of numerical experiments, we list exact one- and two- soliton solutions for the continuous, semi- and full-discrete CSP equation.

{\bf (1). One soliton solution}: the $\tau$-functions   for one
soliton solution of the CSP equation (\ref{CSPE1})--(\ref{CSPE2}) are
\begin{equation}
\label{1-soliton_con1}
f \propto 1+ {\rm i} e^{\theta}\,\quad g \propto 1 + {\rm i} s_1 e^{\theta}\,,
\end{equation}
where $s_1=(1-p_1)/(1+p_1)$, $\theta=p_1y+s/p_1+y_0$. This leads to one-soliton solution in parametric form
\begin{eqnarray}\label{1-soliton_con2}
    u(y,s)= \frac{1}{p_1} \left( \rm{sech} \theta + \rm{sgn} (s_1) \rm{sech} (\theta-\Delta) \right), \\
    v(y,s)= \frac{1}{p_1} \left( \rm{sech} \theta - \rm{sgn} (s_1) \rm{sech} (\theta-\Delta) \right),
\end{eqnarray}
\begin{equation}
\label{1-soliton_con3}
   x=y- \frac{1}{p_1} \left( {\tanh} (\theta) + \tanh (\theta-\Delta) \right)\,.
\end{equation}
where $\exp(-\Delta)=|s_1|$.

For the semi-discrete CSP equation, the $\tau$-functions are
\begin{equation}
\label{1-soliton_semidiscrete1}
f_k \propto 1+ {\rm i} \left(\frac{1+ap_1}{1-ap_1} \right)^k e^{s/p_1+y_0}\,,\ \  g_k \propto 1 + {\rm i} s_1
\left(\frac{1+ap_1}{1-ap_1} \right)^k e^{s/p_1+y_0}\,.
\end{equation}

Finally for the fully discrete CSP equation, the $\tau$-functions are
\begin{equation}
\label{1-soliton_fulldiscretea}
f_{k,l} \propto 1+ {\rm i} \left(\frac{1+ap_1}{1-ap_1} \right)^k
\left(\frac{1+bp^{-1}_1}{1-bp^{-1}_1} \right)^l \,,
\end{equation}

\begin{equation}
\label{1-soliton_fulldiscreteb}
g_{k,l} \propto 1 + {\rm i} s_1
\left(\frac{1+ap_1}{1-ap_1} \right)^k \left(\frac{1+bp^{-1}_1}{1-bp^{-1}_1} \right)^l\,.
\end{equation}

\begin{figure}[htbp]
\centerline{
\includegraphics[scale=0.4]{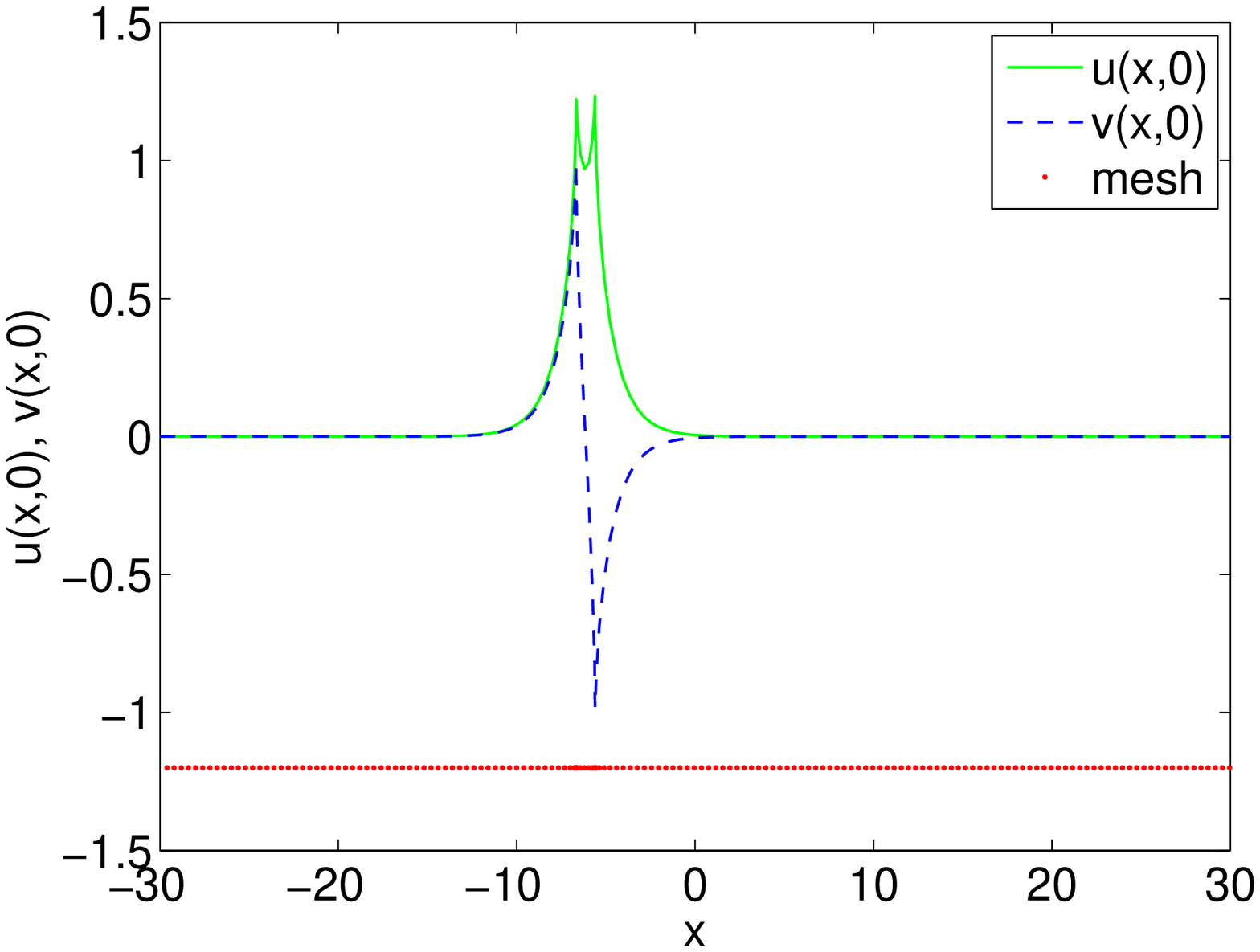}\quad
\includegraphics[scale=0.4]{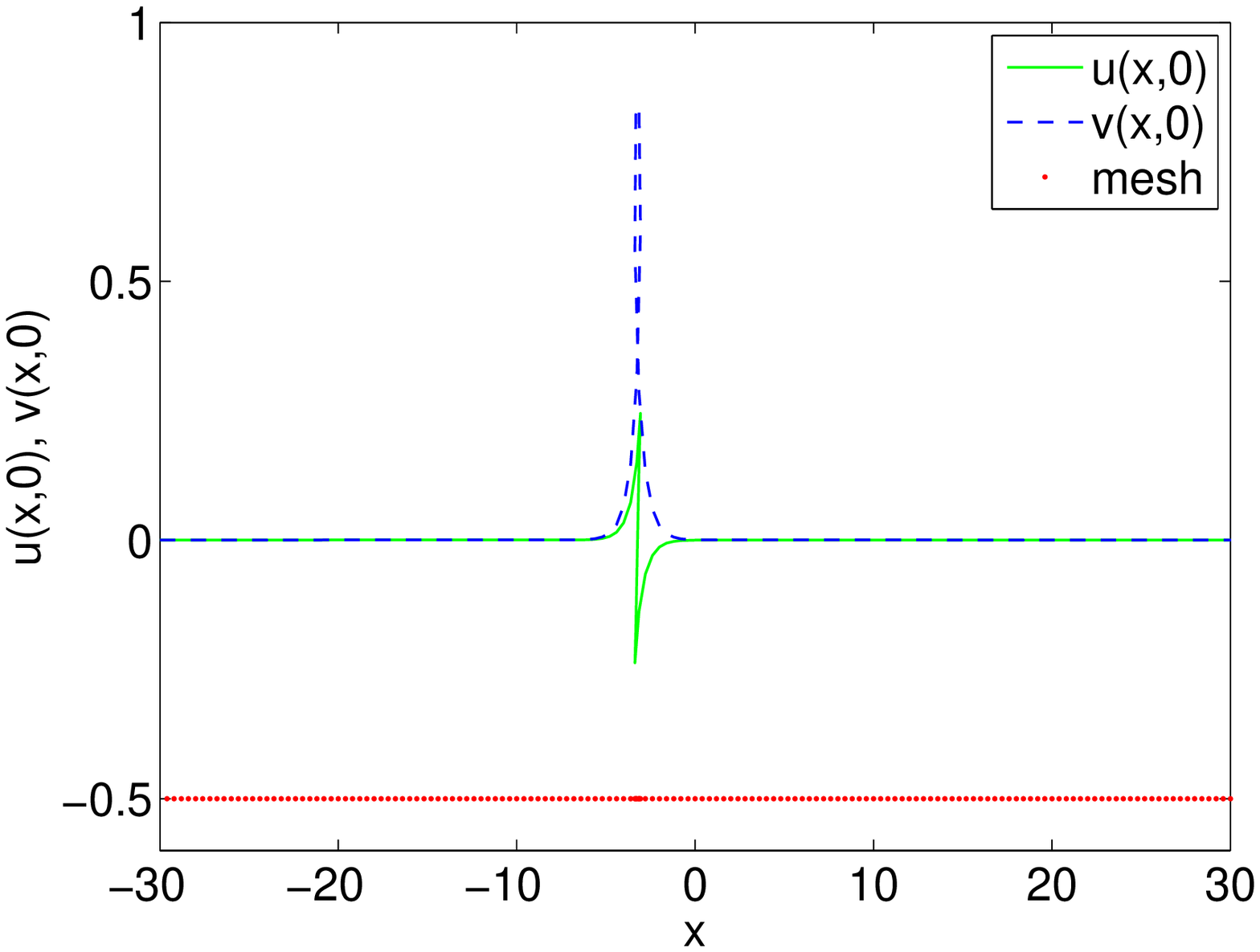}}
\kern-0.315\textwidth \hbox to
\textwidth{\hss(a)\kern15em\hss(b)\kern14em} \kern+0.355\textwidth
\caption{Initial conditions for CSP equation. (a) $p_1=0.9$; (b) $p_1=2.0$.} \label{f:1loopinitial}
\end{figure}

 The initial conditions for one-soliton propagation are taken from (\ref{1-soliton_con1}) with parameters $y_0=0$,  and $p_1=0.9$, $p_1=2.0$. The initial profiles are shown in Fig. \ref{f:1loopinitial}(a) and (b), respectively.
 The simulations are run on a domain $[-40,40]$ with $800$ grid points, thus the average mesh size is $0.1$.

\begin{figure}[htbp]
\centerline{
\includegraphics[scale=0.4]{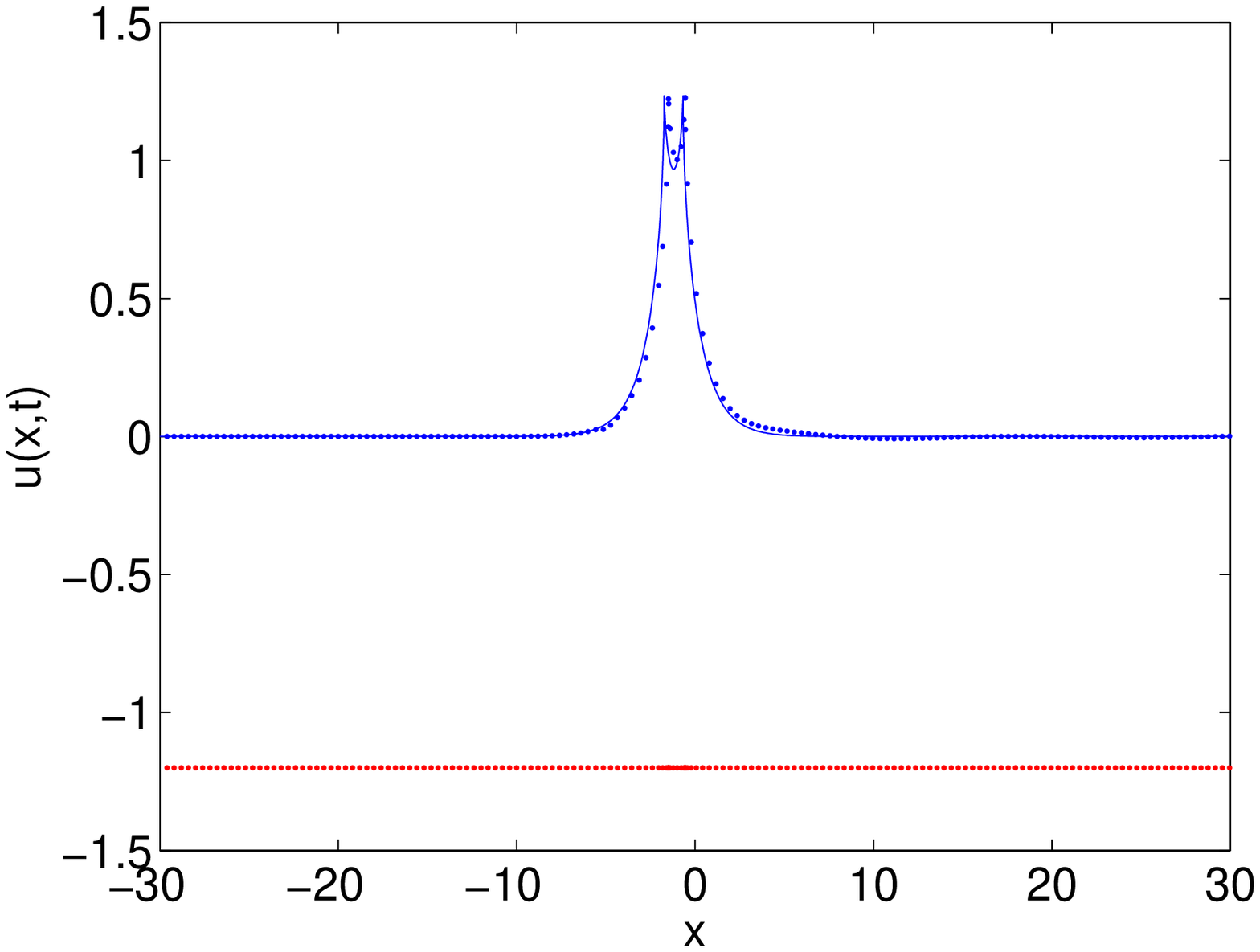}\quad
\includegraphics[scale=0.4]{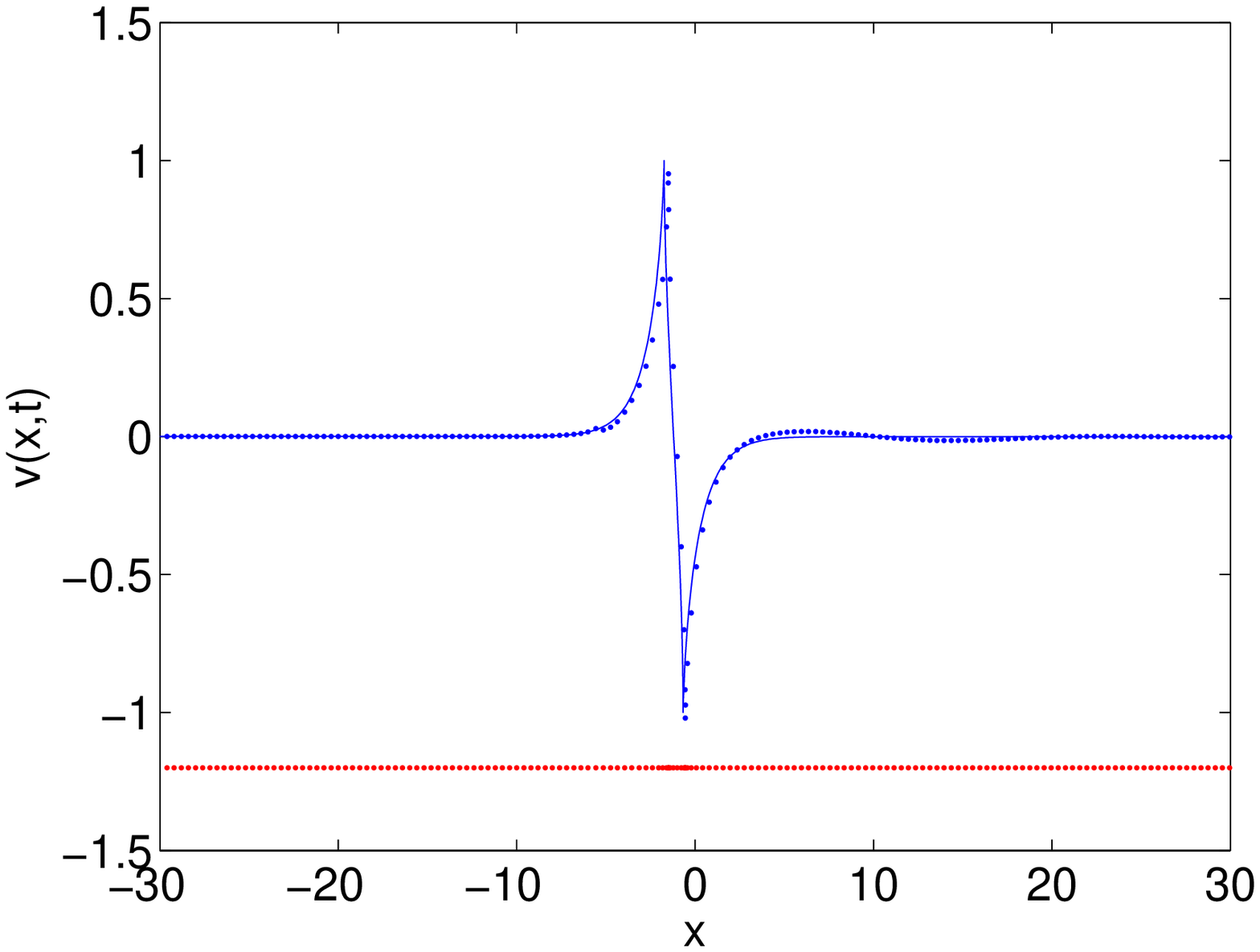}}
\kern-0.315\textwidth \hbox to
\textwidth{\hss(a)\kern14em\hss(b)\kern14em} \kern+0.355\textwidth
\caption{Comparison between numerical and analytical solutions for one-soliton  to the CSP equation with $p_1=0.9$ at $t=4.0$; solid line: analytical solution, blue dot: numerical solution, red dot: self-adaptive mesh;(a) profile of $u$,  (b) profile of $v$.} \label{f:1loopa}
\end{figure}

\begin{figure}[htbp]
\centerline{
\includegraphics[scale=0.4]{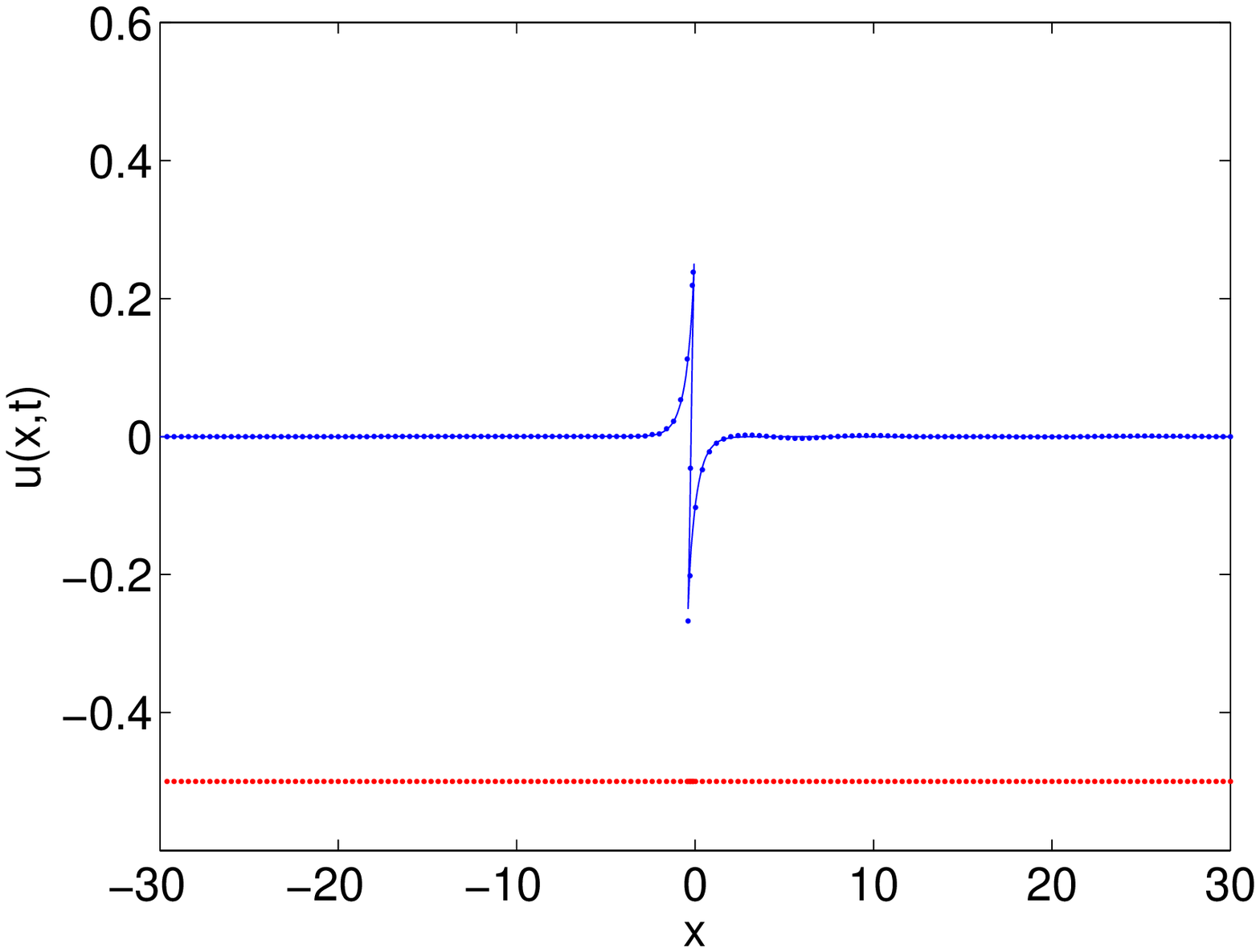}\quad
\includegraphics[scale=0.4]{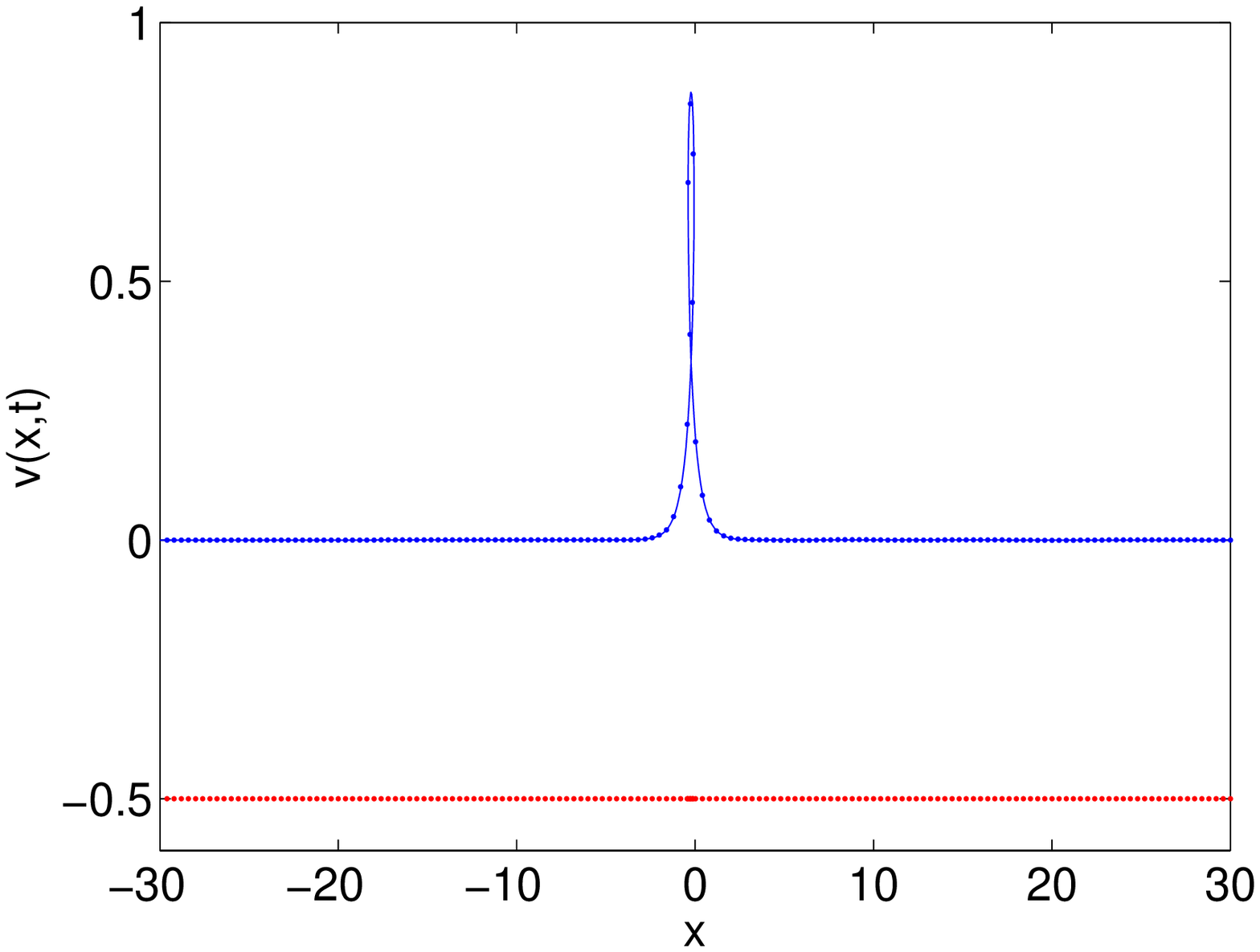}}
\kern-0.315\textwidth \hbox to
\textwidth{\hss(a)\kern15em\hss(b)\kern14em} \kern+0.355\textwidth
\caption{Comparison between numerical and analytical results for one-soliton solution to the CSP equation with $p_1=2.0$ at $t=12.0$; solid line: analytical solution, blue dot: numerical solution, red dot: self-adaptive mesh; (a) profile of $u$,  (b) profile of $v$.} \label{f:1loopb}
\end{figure}

\begin{figure}[htbp]
\centerline{
\includegraphics[scale=0.4]{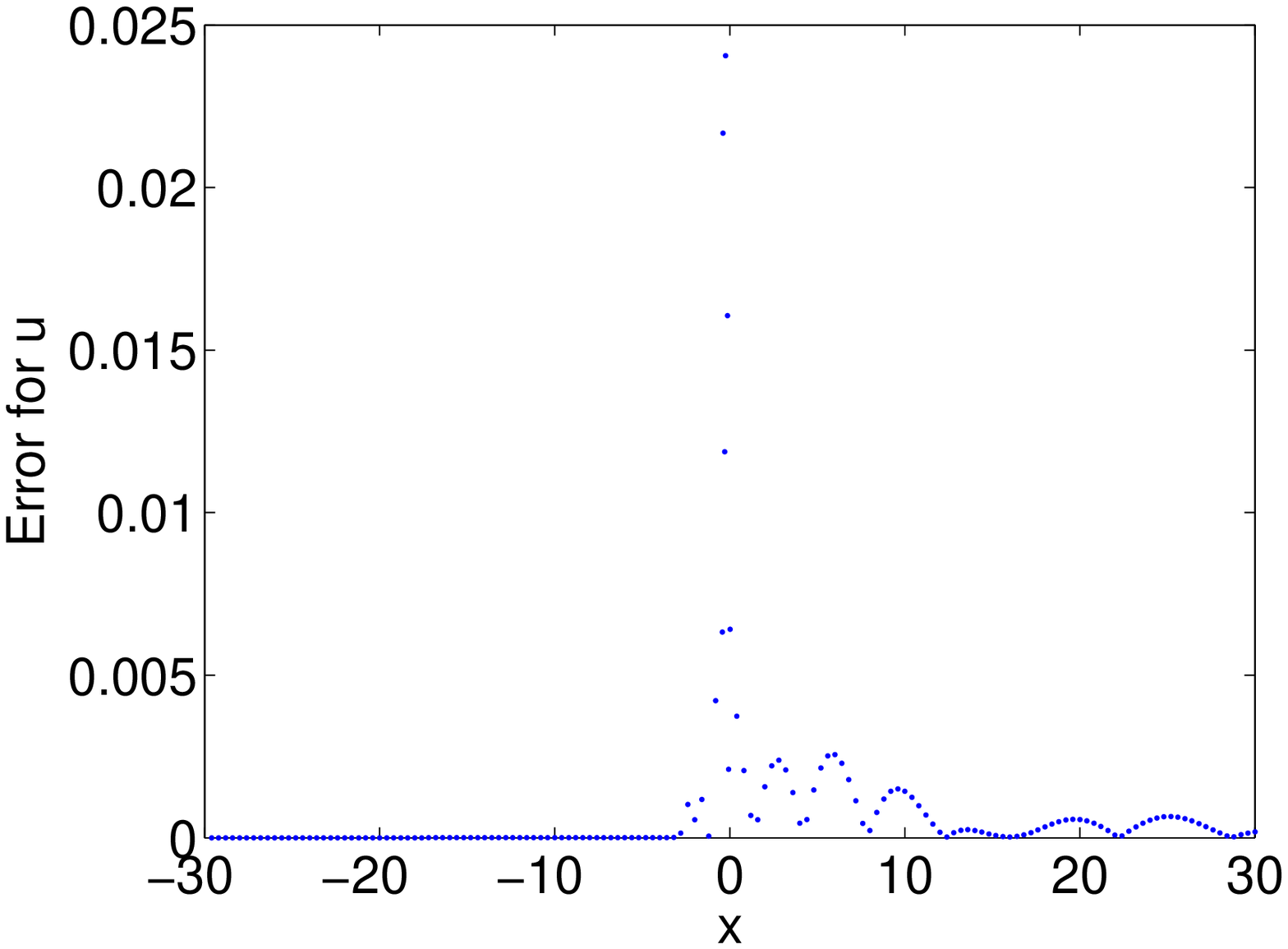}\quad
\includegraphics[scale=0.4]{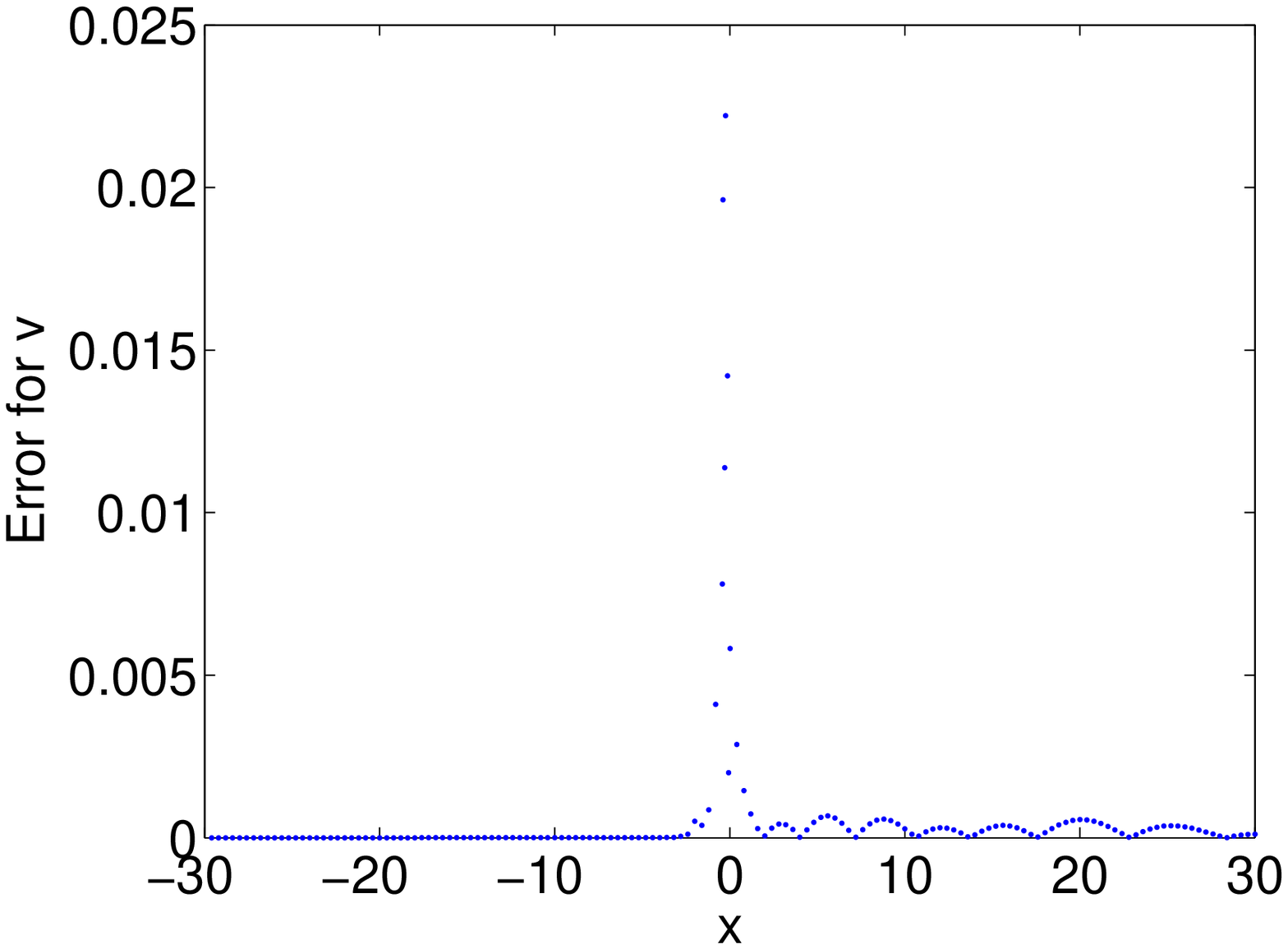}}
\kern-0.315\textwidth \hbox to
\textwidth{\hss(a)\kern15em\hss(b)\kern14em} \kern+0.355\textwidth
\caption{The error between numerical and analytical results for one-soliton solution to the CSP equation with $p_1=2.0$ at $t=12.0$; (a) error in $u$,  (b) error in $v$.} \label{f:1loopc}
\end{figure}

 When $p_1=0.9$, $u$ is symmetric with two-spikes, and $v$ is antisymmetric.
The comparison between the numerical and analytical results is shown in Fig. \ref{f:1loopa}, together with the nonuniform mesh. It can be seen that the non-uniform mesh is dense around the peak points of solitons. Moreover, the most dense part of the non-uniform mesh moves along with the peak point.
When $p_1=2.0$, $u$ is antisymmetric, and $v$ is symmetric with a loop structure. The comparison between the numerical and analytical results is shown in Fig. \ref{f:1loopb}.
The error between the numerical solution and the analytical one is displayed in Fig. \ref{f:1loopc}.

{\bf (2). Two soliton solution}: the $\tau$-functions  for two
soliton solution of the CSP equation (\ref{CSPE1})--(\ref{CSPE2}) are
\begin{eqnarray}
f & \propto & 1 + {\rm i} e^{\theta_1}+ {\rm i} e^{\theta_2}- b_{12}  e^{\theta_1+\theta_2}\,, \label{2-soliton_con1} \\
g & \propto & 1 + {\rm i}s_1 e^{\theta_1}+ {\rm i} s_2 e^{\theta_2}- b'_{12}  e^{\theta_1+\theta_2}\,, \label{2-soliton_con2}
\end{eqnarray}
with $s_i=(1-p_i)/(1+p_i)$, $\theta_i=p_iy+s/p_1+y_{i0}$ $(i=1,2)$, and $b_{12}=(p_1-p_2)^2/(p_1+p_2)^2$,
and $b'_{12}=b_{12}*s_1*s_2$.

For the semi-discrete CSP equation, the $\tau$-functions are
\begin{eqnarray}
f_k & \propto & 1 + {\rm i} \left(\frac{1+ap_1}{1-ap_1} \right)^k  e^{\xi_1}+ {\rm i} \left(\frac{1+ap_2}{1-ap_2} \right)^ke^{\xi_2} \nonumber \\ && - \left(\frac{(1+ap_1)(1+ap_2)}{(1-ap_1)(1-ap_2)} \right)^k b_{12}  e^{\xi_1+\xi_2}\,, \label{2-soliton_semidiscrete1} \\
g_k & \propto & 1 + {\rm i} s_1 \left(\frac{1+ap_1}{1-ap_1} \right)^k  e^{\xi_1}+ {\rm i} s_2 \left(\frac{1+ap_2}{1-ap_2} \right)^ke^{\xi_2} \nonumber \\ && - b'_{12}  \left(\frac{(1+ap_1)(1+ap_2)}{(1-ap_1)(1-ap_2)} \right)^k  e^{\xi_1+\xi_2}\,, \label{2-soliton_semidiscrete2}
\end{eqnarray}
with $\xi_i=s/p_1+y_{i0}$ $(i=1,2)$.

For the fully discrete CSP equation, the $\tau$-functions are
\begin{eqnarray}
f_{k,l} & \propto & 1 + {\rm i} \left(\frac{1+ap_1}{1-ap_1} \right)^k  \left(\frac{1+bp^{-1}_1}{1-bp^{-1}_1} \right)^l + {\rm i} \left(\frac{1+ap_2}{1-ap_2} \right)^k \left(\frac{1+bp^{-1}_2}{1-bp^{-1}_2} \right)^l \nonumber \\ && - b_{12} \left(\frac{(1+ap_1)(1+ap_2)}{(1-ap_1)(1-ap_2)} \right)^k
\left(\frac{(1+bp^{-1}_1)(1+bp^{-1}_2)}{(1-bp^{-1}_1)(1-bp^{-1}_2)} \right)^l \,, \label{2-soliton_fulldiscrete1}
\\
g_{k,l} & \propto & 1 + {\rm i} s_1 \left(\frac{1+ap_1}{1-ap_1} \right)^k \left(\frac{1+bp^{-1}_1}{1-bp^{-1}_1} \right)^l + {\rm i} s_2 \left(\frac{1+ap_2}{1-ap_2} \right)^k \left(\frac{1+bp^{-1}_2}{1-bp^{-1}_2} \right)^l \nonumber \\ && - b'_{12}  \left(\frac{(1+ap_1)(1+ap_2)}{(1-ap_1)(1-ap_2)} \right)^k
\left(\frac{(1+bp^{-1}_1)(1+bp^{-1}_2)}{(1-bp^{-1}_1)(1-bp^{-1}_2)} \right)^l \,. \label{2-soliton_fulldiscrete2}
\end{eqnarray}

\begin{figure}[htbp]
\centerline{
\includegraphics[scale=0.4]{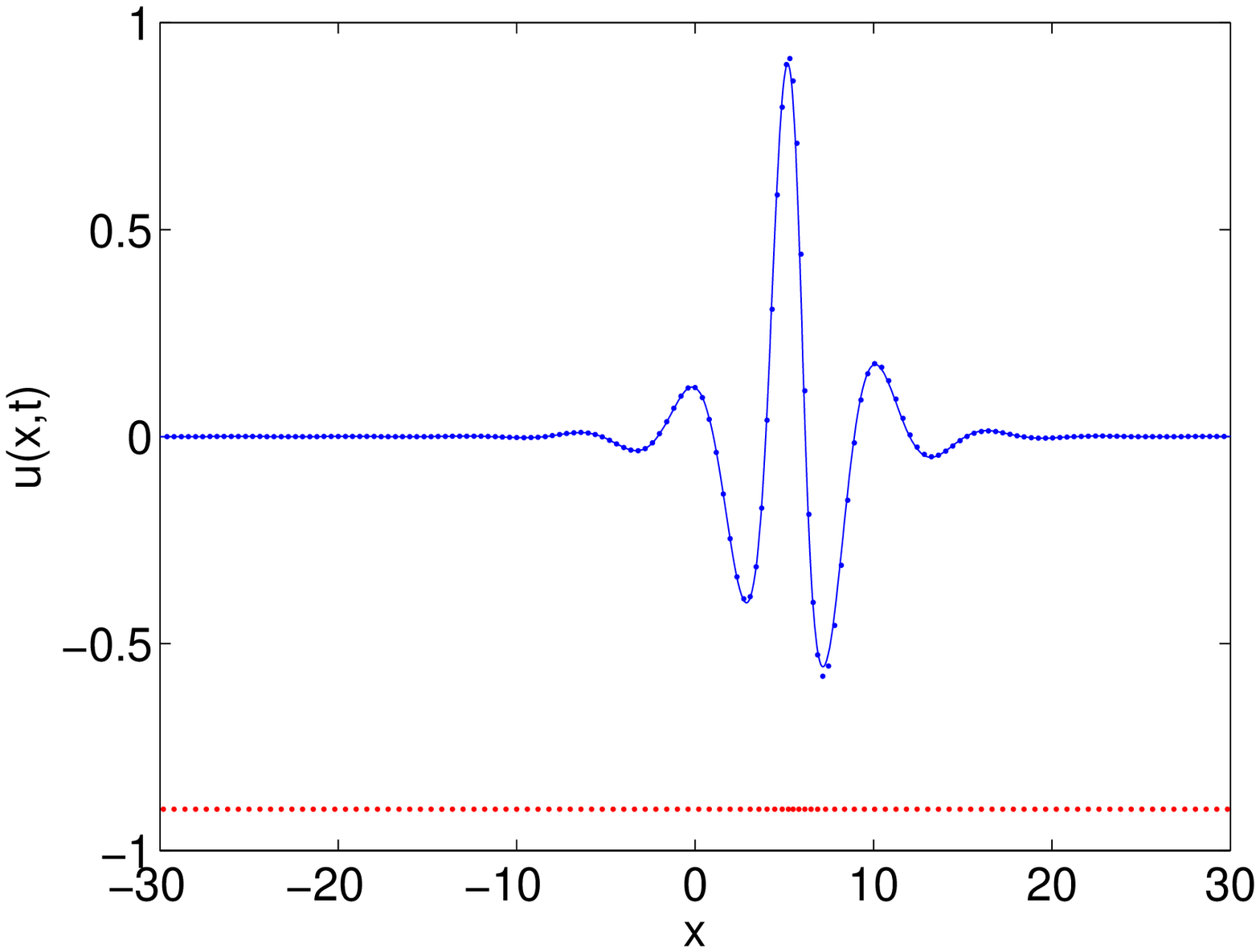}\quad
\includegraphics[scale=0.4]{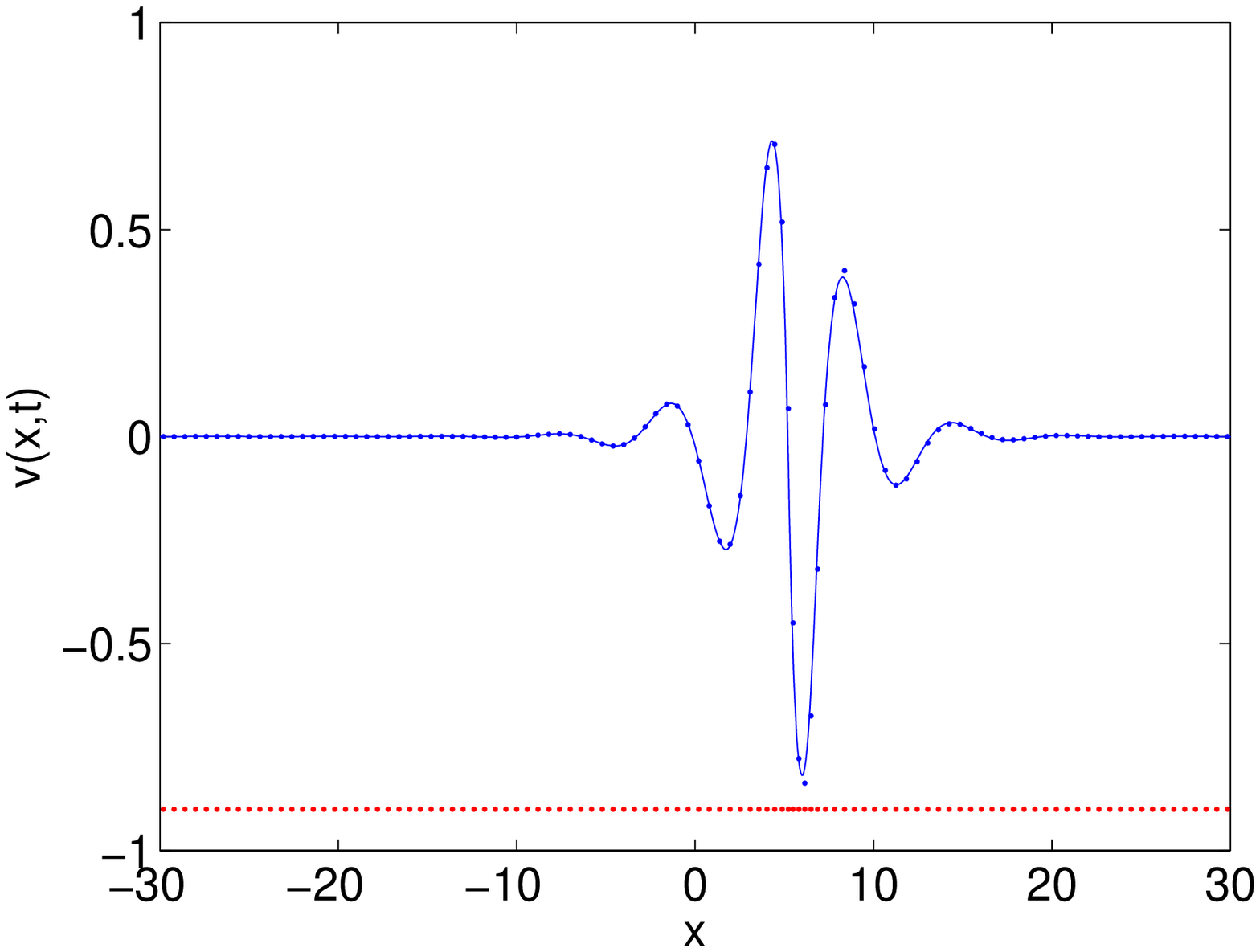}}
\kern-0.315\textwidth \hbox to
\textwidth{\hss(a)\kern15em\hss(b)\kern14em} \kern+0.355\textwidth
\caption{Comparison between numerical and analytical results for breather solution to the CSP equation for $p_1=0.4+{\rm i}$, $p_2=0.4-{\rm i}$ at $t=10$; solid line: analytical solution, dashed line: numerical solution; (a) profile of $u$,  (b) profile of $v$.} \label{f:breather}
\end{figure}

\begin{figure}[htbp]
\centerline{
\includegraphics[scale=0.4]{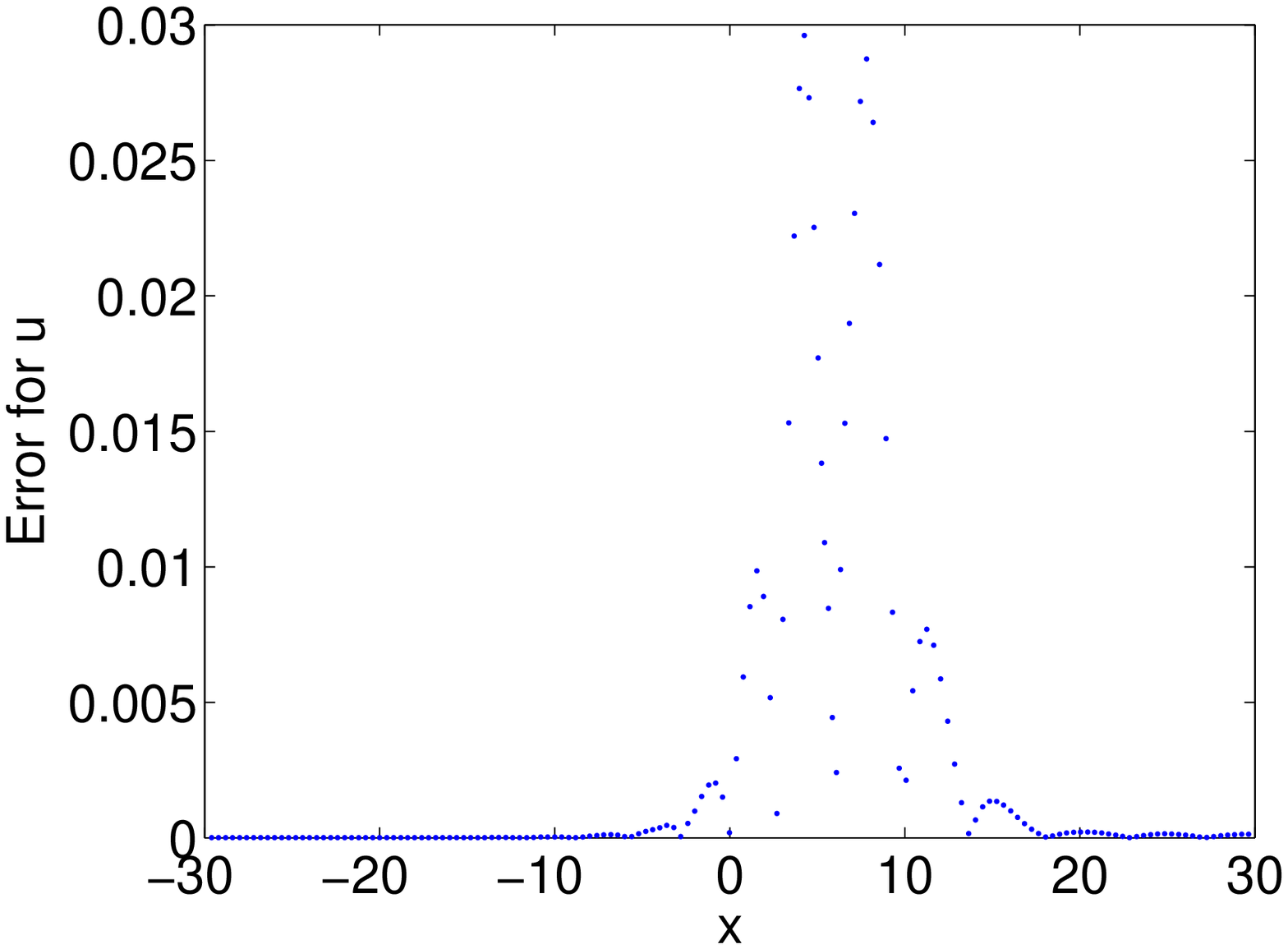}\quad
\includegraphics[scale=0.4]{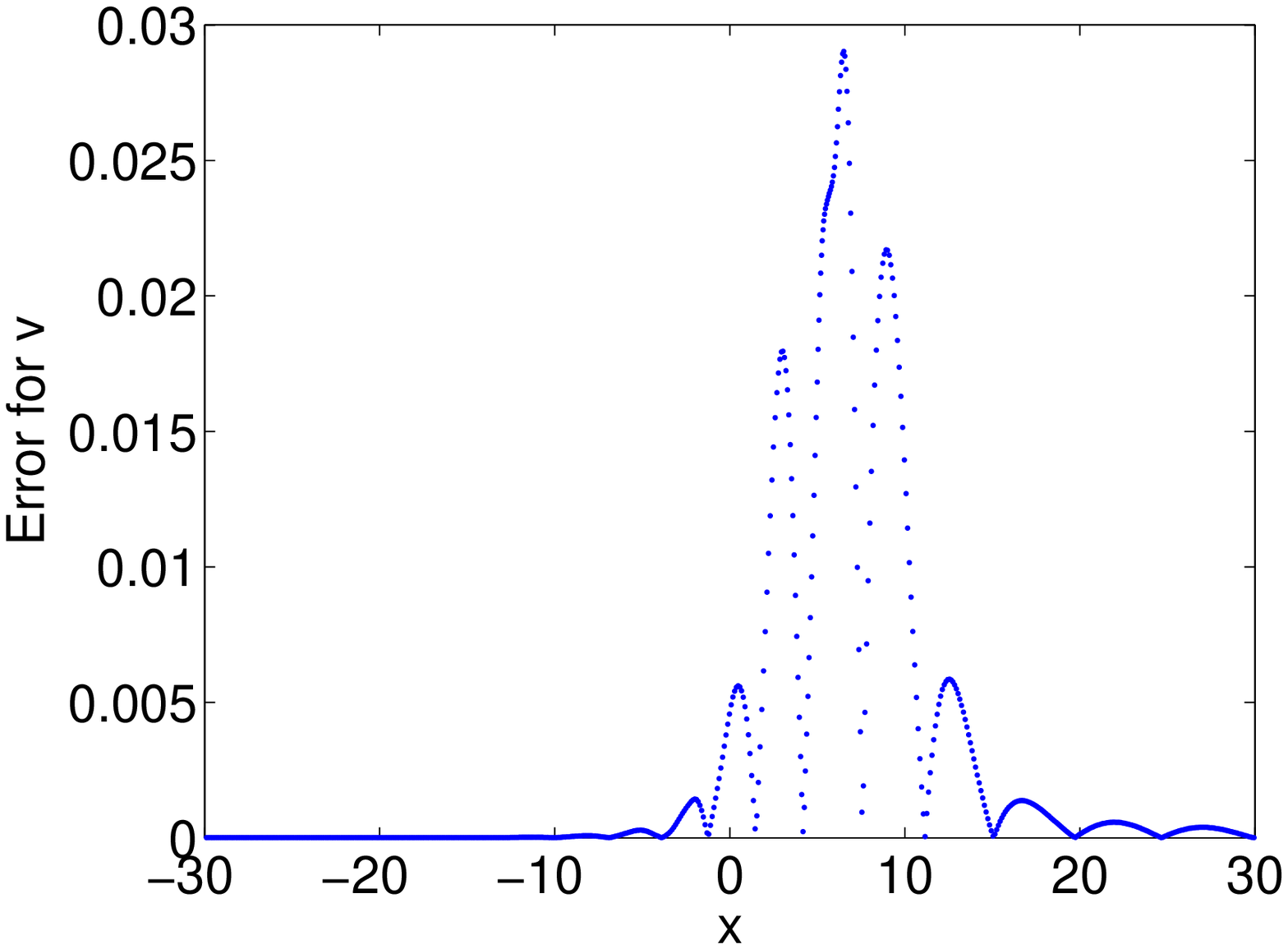}}
\kern-0.315\textwidth \hbox to
\textwidth{\hss(a)\kern15em\hss(b)\kern14em} \kern+0.355\textwidth
\caption{The error between numerical and analytical results for one-breather solution to the CSP equation at $t=10.0$; (a) error in $u$,  (b) error in $v$.} \label{f:1breather}
\end{figure}

As pointed in \cite{CSPE_Feng}, when $p_1$ and $p_2$ are complex conjugate to each other, two-soliton solution becomes a breather solution. Eqs. (\ref{2-soliton_con1})--(\ref{2-soliton_con2}) are used as initial condition with parameters chosen as $p_1=0.4+{\rm i}$, $p_2=0.4-{\rm i}$, $y_{10}=y_{20}=0$. The numerical results at $t=10$ are displayed in Fig. \ref{f:breather} in compared with analytical solution. Here a grid points of $800$ is used on a domain $[-40,40]$, the time step size is taken as $\Delta t= 0.005$. The error between the numerical solution and the analytical one is displayed in Fig. \ref{f:1breather}.  It can be seen that the numerical results are in good agreement with analytical ones.

\section{Conclusions}
In this paper, we proposed integrable semi-discrete and fully discrete analogues of a coupled short pulse equation.
The determinant formulas of $N$-soliton solutions for the semi-discrete and fully discrete analogues of the CSP equations are also presented.
In the continuous limit, the fully discrete CSP equation converges to the semi-discrete CSP equation,
then further converges to the continuous CSP equation.

In a series of papers by one of the authors, we have constructed integrable discretizations for a class of soliton equations with hodograph transformation, and successfully used them as self-adaptive moving mesh methods for the Camassa-Holm equation \cite{dCH,dCHcom} and the short pulse equation \cite{CSPE_Feng}. Based on the semi-discrete CSP equation (\ref{CSPE1})--(\ref{CSPE2}), a self-adaptive moving mesh method is constructed and used for the numerical simulations of the CSP equation. It should be pointed out that the feature of self-adaptivity of the mesh is due to the hodograph transformation. In other words, the hodograph transformation converts the uniform and time-independent mesh into a non-uniform and time-dependent mesh. It is a further topic to seek for such kind of self-adaptive moving method when the hodograph transformation is not present.
The numerical results confirms that it is an excellent scheme due to its nature of integrability and self-adaptivity of the mesh. This is our first time to extend this superior numerical method to a coupled system.

\section*{Acknowledgments}

This work is partially supported by the National Natural Science Foundation of China (Nos. 11428102, 11075055, 11275072).

\appendix

\section{Proof of Proposition 1}
\begin{proof}
 For simplicity, we introduce a convenient notation
\begin{equation}
\left\vert 0_k, 1_k,  \cdots, N-1_k  \right\vert
=\left\vert \begin{array}{cccc}
  \phi_n^{(1)}(k) & \phi_{n+1}^{(1)}(k) & \cdots & \phi_{n+N-1}^{(1)}(k) \\
  \phi_n^{(2)}(k) & \phi_{n+1}^{(2)}(k) & \cdots & \phi_{n+N-1}^{(2)}(k) \\
  \cdots & \cdots & \cdots & \cdots \\
  \phi_n^{(N)}(k) & \phi_{n+1}^{(N)}(k) & \cdots & \phi_{n+N-1}^{(N)}(k) \\
  \end{array}\right\vert,
\end{equation}
\begin{equation}
 \left\vert 0'_k, 1'_k, \cdots, N-1'_k  \right\vert
  =\left\vert \begin{array}{cccc}
  \psi_n^{(1)}(k) & \psi_{n+1}^{(1)}(k) & \cdots & \psi_{n+N-1}^{(1)}(k) \\
  \psi_n^{(2)}(k) & \psi_{n+1}^{(2)}(k) & \cdots & \psi_{n+N-1}^{(2)}(k) \\
  \cdots & \cdots & \cdots & \cdots \\
  \psi_n^{(N)}(k) & \psi_{n+1}^{(N)}(k) & \cdots & \psi_{n+N-1}^{(N)}(k) \\
  \end{array}\right\vert.
\end{equation}
Since
\begin{equation}
\partial_{x_{-1}}\phi^{(i)}_n(k)=\phi^{(i)}_{n-1}(k),\ \ \phi^{(i)}_{n}(k+1)-\phi^{(i)}_{n}(k)=a \phi^{(i)}_{n+1}(k+1),
\end{equation}

\begin{equation}
\partial_{x_{-1}}\psi^{(i)}_n(k)=\psi^{(i)}_{n-1}(k),\ \ \psi^{(i)}_{n}(k+1)-\psi^{(i)}_{n}(k)=a \psi^{(i)}_{n+1}(k+1),
\end{equation}

\begin{equation}
\phi^{(i)}_n(k) - \psi^{(i)}_{n}(k)=\phi^{(i)}_{n+1}(k),
\end{equation}
we can verify the following relations
\begin{equation}
\partial_{x_{-1}}\tau_n(k) = \left\vert -1_k, 1_k, \cdots, N-2_k, N-1_k  \right\vert\,,
\end{equation}

\begin{eqnarray}
 \tau_{n+1}(k{+}1) &=& \left\vert 1_{k}, 2_{k}, \cdots,N-2_{k},  N-1_{k}, N_{k+1}  \right\vert \nonumber \\
 & = &\frac{1}{a}\left\vert 1_{k}, 2_{k}, \cdots,N-2_{k},  N-1_{k}, N-1_{k+1}  \right\vert\,,
\end{eqnarray}

\begin{eqnarray}
   \tau_n(k{+}1) &=& \left\vert 0_{k+1}, 1_{k+1}, \cdots, N-2_{k+1}, N-1_{k+1}  \right\vert \nonumber \\
   &= &\left\vert 0_{k}, 1_{k},  \cdots,  N-2_{k}, N-1_{k+1}  \right\vert \,,
   \label{temp1}
\end{eqnarray}

\begin{eqnarray}
 \hspace{-1.5cm}  \partial_{x_{-1}} \tau_n(k+1) &=& \left\vert -1_{k}, 1_{k},  \cdots,  N-2_{k}, N-1_{k+1}  \right\vert
    + \left\vert 0_{k}, 1_{k},  \cdots,  N-2_{k}, N-2_{k+1}  \right\vert \nonumber \\
\hspace{-1.5cm}   & = & \left\vert -1_{k}, 1_{k},  \cdots,  N-2_{k}, N-1_{k+1}  \right\vert
    + a \left\vert 0_{k}, 1_{k},  \cdots,  N-2_{k}, N-1_{k+1}  \right\vert\,.
    \label{temp2}
\end{eqnarray}
Combining (\ref{temp1}) with (\ref{temp2}), we have
\begin{equation}
\left(\frac{1}{a}\partial_{x_{-1}}-1\right)\tau_n(k+1)=\frac{1}{a}\left\vert -1_{k}, 1_{k}, \cdots,  N-2_{k}, N-1_{k+1}  \right\vert\,.
\end{equation}
Therefore, the Pl\"{u}cker relation for determinants
\begin{eqnarray}
&& \left\vert 0_k, 1_k, \cdots, N-2_k, N-1_k  \right\vert \left\vert -1_{k}, 1_{k},  \cdots,  N-2_{k}, N-1_{k+1}  \right\vert \nonumber \\
&& -\left\vert 0_{k}, 1_{k},  \cdots,  N-2_{k}, N-1_{k+1}  \right\vert \left\vert -1_k, 1_k,  \cdots,N-2_k, N-1_k  \right\vert \nonumber \\
&& +\left\vert 1_{k}, \cdots,N-2_{k},  N-1_{k}, N-1_{k+1}  \right\vert \left\vert -1_k, 0_k, 1_k, \cdots, N-2_k  \right\vert=0
\end{eqnarray}
gives
\begin{equation}
\hspace{-2.2cm}\left(\frac{1}{a}\partial_{x_{-1}}-1\right) \tau_n(k+1)\times \tau_n(k)-\frac{1}{a} \tau_n(k+1) \times \partial_{x_{-1}} \tau_n(k)
 +\tau_{n+1}(k+1)\tau_{n-1}(k)= 0\,,
\end{equation}
which is nothing but the bilinear equation (\ref{BK2DTL1}). Eq. (\ref{BK2DTL2}) can be proved in the same way.

Now we proceed to the proof of Eq. (\ref{BT}). Similarly we can verify the following relations
\begin{eqnarray}
 \tau_n(k) &=& \left\vert 0_k, 1_k, \cdots, N-2_k, N-1_k  \right\vert \nonumber \\
 & = &\left\vert 0'_k, 1'_k, \cdots, N-2'_k, N-1_k  \right\vert\,,
\end{eqnarray}

\begin{eqnarray}
 \tau_{n+1}(k) &= & \left\vert 1'_k, 2'_k, \cdots, N-1'_k, N_k  \right\vert \nonumber \\
 & = &\left\vert 1'_k, 2'_k, \cdots, N-1'_k, N-1_k  \right\vert\,,
\end{eqnarray}

\begin{equation}
\left(\partial_{x_{-1}}-1\right)\tau_n(k)=\left\vert -1'_k, 1'_k, \cdots, N-2'_k, N-1_k  \right\vert\,.
\end{equation}

Therefore the Pl\"{u}cker relation for determinants:
\begin{eqnarray}
&& \left\vert -1'_k, 1'_k,  \cdots, N-2'_k, N-1_k  \right\vert \left\vert 0'_k, 1'_k,  \cdots, N-2'_k, N-1'_k  \right\vert \nonumber \\
&& - \left\vert 0'_k, 1'_k,  \cdots, N-2'_k, N-1_k  \right\vert  \left\vert -1'_k, 1'_k,  \cdots, N-2'_k, N-1'_k  \right\vert  \nonumber \\
&& + \left\vert -1'_k, 0'_k, 1'_k, \cdots, N-2'_k  \right\vert \left\vert 1'_k,  \cdots, N-1'_k, N-1_k  \right\vert =0\,,
\end{eqnarray}
gives
\begin{eqnarray}
&& (\partial_{x_{-1}}-1)\tau_n(k)\times\tau'_n(k) - \tau_n(k)\times \partial_{x_{-1}}\tau'_n(k) + \tau_{n+1}(k)\tau'_{n-1}(k) =0\,.
\end{eqnarray}
Therefore, Eq.(\ref{BT}) is proved.
\end{proof}
\section{Proof of Theorem 1}
\begin{proof}
By putting $s=2x_{-1}$, $\tau_0(k)=f_k$, $\tau_1(k)=\bar{f}_k$, (\ref{BK2DTL1}) can be converted into
\begin{eqnarray}
  \left(\frac{2}{a}D_s -1\right)f_{k+1}\cdot f_k =-\bar{f}_{k+1}\bar{f}_k\,,  \\
 \left(\frac{2}{a}D_s -1\right)\bar{f}_{k+1} \cdot \bar{f}_k =-f_{k+1}f_k\,,
\end{eqnarray}
while by putting $\tau'_0(k)=g_k$, $\tau'_1(k)=\bar{g}_k$, (\ref{BK2DTL2}) can be converted into
\begin{eqnarray}
  \left(\frac{2}{a}D_s -1\right)g_{k+1}\cdot g_k =-\bar{g}_{k+1}\bar{g}_k\,,  \\
 \left(\frac{2}{a}D_s -1\right)\bar{g}_{k+1} \cdot \bar{g}_k =-g_{k+1}g_k\,.
\end{eqnarray}

Furthermore, the above bilinear equations can be rewritten as the following logarithmic derivatives
\begin{eqnarray}
&&\frac{2}{a}
\left(\ln \frac{f_{k+1}}{f_k}\right)_s
-1
= - \frac{\bar{f}_{k+1}\bar{f}_k}{f_{k+1}f_k}\,,\label{2DDTD1} \\
&&\frac{2}{a}
\left(\ln \frac{\bar{f}_{k+1}}{\bar{f}_k}\right)_s
-1
= -\frac{{f}_{k+1}{f}_k}{\bar{f}_{k+1}\bar{f}_k}\,,\label{2DDTD2}
\end{eqnarray}
\begin{eqnarray}
&&\frac{2}{a}
\left(\ln \frac{g_{k+1}}{g_k}\right)_s
-1
= - \frac{\bar{g}_{k+1}\bar{g}_k}{g_{k+1}g_k}\,,\label{2DDTD3}\\
&&\frac{2}{a}
\left(\ln \frac{\bar{g}_{k+1}}{\bar{g}_k}\right)_s
-1
= -\frac{{g}_{k+1}{g}_k}{\bar{g}_{k+1}\bar{g}_k}\,. \label{2DDTD4}
\end{eqnarray}

Introducing two intermediate variable transformations $$\sigma_k(s)=2{\rm i}\ln\left(\frac{\bar{f}_k(s)}{f_k(s)}\right), \quad
\sigma'_k(s)=2{\rm i}\ln\left(\frac{\bar{g}_k(s)}{g_k(s)}\right)\,,$$
one arrives at a pair of semi-discrete sine-Gordon equations
\begin{equation}
\frac{1}{2a} \left( \sigma_{k+1}-\sigma_{k} \right)_s=\sin
\left(\frac{\sigma_{k+1}+\sigma_k}{2}\right)\,,
\label{semi-SG1}
\end{equation}

\begin{equation}
\frac{1}{2a}  \left( \sigma'_{k+1}-\sigma'_{k} \right)_s=\sin
\left(\frac{\sigma'_{k+1}+\sigma'_k}{2}\right)\,.
\label{semi-SG2}
\end{equation}
It then follows that
\begin{equation}
\left(\cos \left(\frac{\sigma_{k+1}+\sigma_k}{2}\right)\right)_s
=-\frac{1}{4a} \left( \left(\sigma_{k+1,s}\right)^2-\left(\sigma_{k,s}\right)^2\right)\,,
\end{equation}

\begin{equation}
\left(\cos \left(\frac{\sigma'_{k+1}+\sigma'_k}{2}\right)\right)_s
=-\frac{1}{4a} \left( \left(\sigma'_{k+1,s}\right)^2-\left(\sigma'_{k,s}\right)^2\right)\,,
\end{equation}
where $\sigma_{k,s}$ denoted the derivative of $\sigma_{k}$ with respective to $s$.
\par
Next, we introduce dependent variable transformations
\begin{equation}
u_k=\frac 12  \left( \sigma_{k,s} + \sigma'_{k,s}\right)  = {\rm i}\ln\left(\frac{\bar{f}_k \bar{g}_k}{f_k g_k}\right)_s\,,
\end{equation}

\begin{equation}
v_k=\frac 12  \left( \sigma_{k,s} - \sigma'_{k,s}\right)  = {\rm i}\ln\left(\frac{\bar{f}_k g_k}{f_k \bar{g}_k}\right)_s\,,
\end{equation}
and discrete hodograph transformation
\begin{equation}
x_k=2ka- \left( \ln(f_k \bar{f}_k g_k \bar{g}_k)\right)_s \,.
\end{equation}

Then the nonuniform mesh can be derived as
\begin{eqnarray}
 && \delta_k = x_{k+1}-x_k \nonumber \\
   &=& 2a-  \left(\ln \frac{f_{k+1}\bar{f}_{k+1}g_{k+1}\bar{g}_{k+1}}
   {f_{k}\bar{f}_{k}g_{k}\bar{g}_{k}}\right)_s \nonumber \\
   &=& \frac{a}{2}  \left(\frac{{f}_{k+1}{f}_k}{\bar{f}_{k+1}\bar{f}_k}+\frac{\bar{f}_{k+1}\bar{f}_k}{f_{k+1}f_k} +
   \frac{{g}_{k+1}{g}_k}{\bar{g}_{k+1}\bar{g}_k} + \frac{\bar{g}_{k+1}\bar{g}_k}{g_{k+1}g_k}\right) \nonumber \\
&=& a  \left( \cos \left(\frac{\sigma_{k+1}+\sigma_k}{2}\right) +
\cos \left(\frac{\sigma'_{k+1}+\sigma'_k}{2}\right) \right)\,.
\end{eqnarray}


Taking the derivative with respect to $s$ results in
\begin{eqnarray}
\label{sd-CSP3b}
  \nonumber \frac{d \delta_k}{d\,s} &=& a   \left( \cos \left(\frac{\sigma_{k+1}+\sigma_k}{2}\right)_s +
\cos \left(\frac{\sigma'_{k+1}+\sigma'_k}{2}\right)_s \right)  \\
   &=&  -\frac 12 \left(u^2_{k+1}-u^2_k + v^2_{k+1}-v^2_k \right) \,.
\end{eqnarray}

Furthermore, assuming
$$
p_k=\sec \left( \frac{\sigma_{k} + \sigma_{k+1}+ \sigma'_{k}+\sigma'_{k+1}}{4} \right)\,, \quad
q_k=\sec \left( \frac{\sigma_{k} + \sigma_{k+1}- \sigma'_{k}-\sigma'_{k+1}}{4} \right)\,,
$$
we have
\begin{equation}
  \delta_k=\frac{2a}{p_k q_k}\,,
\end{equation}
\begin{eqnarray}
  \frac{d p^{-1}_k}{d\,s}&=& \frac {d} {ds} \cos \left( \frac{\sigma_{k} + \sigma_{k+1}+ \sigma'_{k}+\sigma'_{k+1}}{4} \right) \nonumber \\
   &=& -\sin  \left( \frac{\sigma_{k} + \sigma_{k+1}+ \sigma'_{k}+\sigma'_{k+1}}{4} \right) \frac{u_k+u_{k+1}}{2}
   \nonumber \\
   &=& -\frac {q_k }{2} \left( \sin  \left( \frac{\sigma_{k} + \sigma_{k+1}}{2}\right) -\sin  \left(  \frac {\sigma'_{k}+\sigma'_{k+1}}{2} \right) \right) \frac{u_k+u_{k+1}}{2} \nonumber \\
   &=&  -\frac {q_k }{2} \frac{1}{2a}  \left( \sigma_{k+1}-\sigma_{k} + \sigma'_{k+1}-\sigma'_{k} \right)_s  \frac{u_k+u_{k+1}}{2}  \nonumber \\
   &=& -\frac{q_k}{4a} \left(u^2_{k+1}-u^2_k \right) \,,
   \label{pk1}
\end{eqnarray}
and similarly
\begin{equation}    \label{qk1}
\frac{d q^{-1}_k}{d\,s} = -\frac{p_k}{4a} \left(v^2_{k+1}-v^2_k \right)\,.
\end{equation}
Thus, in turn, Eqs. (\ref{pk1}) and (\ref{qk1})  become
\begin{equation}
    \frac{d p^{2}_k}{d\,s} = p^{2}_k \frac{u^2_{k+1}-u^2_k}{\delta_k}\,,
\end{equation}
and
\begin{equation}
    \frac{d q^{2}_k}{d\,s} = q^{2}_k \frac{v^2_{k+1}-v^2_k}{\delta_k}\,,
\end{equation}
respectively.

On the other hand, with the help of trigonometric identity, $p^{2}_k$ can be expressed as
\begin{eqnarray}
  p^{2}_k &=&  1+ \tan^2 \left( \frac{\sigma_{k} + \sigma_{k+1}+ \sigma'_{k}+\sigma'_{k+1}}{4} \right) \nonumber \\
   &=& 1+ p^{2}_k \sin^2 \left( \frac{\sigma_{k} + \sigma_{k+1}+ \sigma'_{k}+\sigma'_{k+1}}{4} \nonumber \right) \\
   &=& 1+  \frac{p^{2}_k q^{2}_k}{4a^2} \left( u^2_{k+1}-u^2_k \right) \nonumber \\
   &= & 1+ \left(\frac{u_{k+1}-u_k}{\delta_k}\right)^2 \,,
\end{eqnarray}
and similarly
\begin{equation}
    q^{2}_k = 1+  \left( \frac{v_{k+1}-v_k}{\delta_k}\right)^2  \,.
\end{equation}
Therefore, we finally have
\begin{equation} \label{sd_CSP1b}
    \frac{d} {ds} \left( \frac{u_{k+1}-u_k}{\delta_k}\right) = \left( 1+ \left( \frac{u_{k+1}-u_k}{\delta_k}\right)^2  \right) \frac{u_{k+1}+u_k}{2}\,,
\end{equation}

\begin{equation} \label{sd_CSP2b}
    \frac{d} {ds} \left( \frac{v_{k+1}-v_k}{\delta_k}\right) = \left( 1+ \left( \frac{v_{k+1}-v_k}{\delta_k}\right)^2  \right) \frac{v_{k+1}+v_k}{2}\,.
\end{equation}
Substituting (\ref{sd-CSP3b}) into  (\ref{sd_CSP1b})--(\ref{sd_CSP2b}), one arrives at
 (\ref{sd_CSP-1})--(\ref{sd_CSP-2}), the first two equations of the semi-discrete CSP equation.
%
\end{proof}
\section{Proof of Theorem 2}
 \begin{proof}
The bilinear equations (\ref{Backlund1})--(\ref{Backlund2b}) imply the following bilinear equations
\begin{eqnarray}
&&\left(\frac{2}{a}D_s-1\right)f_{k+1,l}\cdot f_{k,l}
+\bar{f}_{k+1,l}\bar{f}_{k,l}=0\,,\\
&&\left(\frac{2}{a}D_s-1\right)\bar{f}_{k+1,l}\cdot \bar{f}_{k,l}
+{f}_{k+1,l}{f}_{k,l}=0\,,\\
&&\left(\frac{2}{a}D_s-1\right)g_{k+1,l}\cdot g_{k,l}
+\bar{g}_{k+1,l}\bar{g}_{k,l}=0\,,\\
&&\left(\frac{2}{a}D_s-1\right)\bar{g}_{k+1,l}\cdot \bar{g}_{k,l}
+{g}_{k+1,l}{g}_{k,l}=0\,,\\
&&(2bD_s-1)f_{k,l+1}\cdot \bar{f}_{k,l}
+f_{k,l}\bar{f}_{k,l+1}=0\,,\\
&&(2bD_s-1)\bar{f}_{k,l+1}\cdot f_{k,l}
+\bar{f}_{k,l}f_{k,l+1}=0\,, \\
&&(2bD_s-1)g_{k,l+1}\cdot \bar{g}_{k,l}
+g_{k,l}\bar{g}_{k,l+1}=0\,,\\
&&(2bD_s-1)\bar{g}_{k,l+1}\cdot g_{k,l}
+\bar{g}_{k,l}g_{k,l+1}=0\,,
\end{eqnarray}
which can be rewritten by logarithmic derivatives as
\begin{eqnarray}
&&\frac{2}{a}\left(\ln\frac{f_{k+1,l}}{f_{k,l}}\right)_s=1
-\frac{\bar{f}_{k+1,l}\bar{f}_{k,l}}{f_{k+1,l}f_{k,l}}\,,\\
&&\frac{2}{a}\left(\ln\frac{\bar{f}_{k+1,l}}{\bar{f}_{k,l}}\right)_s=1
-\frac{f_{k+1,l}f_{k,l}}{\bar{f}_{k+1,l}\bar{f}_{k,l}}\,,\\
&&\frac{2}{a}\left(\ln\frac{g_{k+1,l}}{g_{k,l}}\right)_s=1
-\frac{\bar{g}_{k+1,l}\bar{g}_{k,l}}{g_{k+1,l}g_{k,l}}\,,\\
&&\frac{2}{a}\left(\ln\frac{\bar{g}_{k+1,l}}{\bar{g}_{k,l}}\right)_s=1
-\frac{g_{k+1,l}g_{k,l}}{\bar{g}_{k+1,l}\bar{g}_{k,l}}\,,\\
&&2b\left(\ln\frac{f_{k,l+1}}{\bar{f}_{k,l}}\right)_s=1
-\frac{f_{k,l}\bar{f}_{k,l+1}}{f_{k,l+1}\bar{f}_{k,l}}\,,\\
&&2b\left(\ln\frac{\bar{f}_{k,l+1}}{f_{k,l}}\right)_s=1
-\frac{\bar{f}_{k,l}f_{k,l+1}}{\bar{f}_{k,l+1}f_{k,l}}\,\\
&&2b\left(\ln\frac{g_{k,l+1}}{\bar{g}_{k,l}}\right)_s=1
-\frac{g_{k,l}\bar{g}_{k,l+1}}{g_{k,l+1}\bar{g}_{k,l}}\,,\\
&&2b\left(\ln\frac{\bar{g}_{k,l+1}}{g_{k,l}}\right)_s=1
-\frac{\bar{g}_{k,l}g_{k,l+1}}{\bar{g}_{k,l+1}g_{k,l}}\,.
\end{eqnarray}


Based on the dependent variable transformation (\ref{full_vartrf}) and discrete hodograph
transformation (\ref{full_hodograph1}), we can verify the following relations
\begin{eqnarray}
\label{fcsp-01} &&{u_{k + 1, \,l}} - {u_{k, \,l}}={\frac {{\rm i}a}{2}}
\,\Big({\frac {{\bar{f}_{k + 1, \,l}}\,{\bar{f}_{k, \,l}}}{{f_{k + 1, \,l}}\,{f_{k, \,l}}}}  +
{ \frac {{\bar{g}_{k + 1, \,l}}\,{\bar{g}_{k
, \,l}}}{{g_{k + 1, \,l}}\,{g_{k, \,l}}}}  - {
\frac {{f_{k + 1, \,l}}\,{f_{k, \,l}}}{{\bar{f}_{k + 1, \,l}}
\,{\bar{f}_{k, \,l}}}}  - { \frac {{g_{k + 1, \,
l}}\,{g_{k, \,l}}}{{\bar{g}_{k + 1, \,l}}\,{\bar{g}_{k,
\,l}}}} \Big), \\
\label{fcsp-02} && {u_{k, \,l + 1}} + {u_{k, \,l}}={
{ \frac {\rm i}{2b}} \,\Big({ \frac {{f_{k,
\,l}}\,{\bar{f}_{k, \,l + 1}}}{{f_{k, \,l + 1}}\,{\bar{f}
_{k, \,l}}}}  + { \frac {{g_{k, \,l}}\,{\bar{g}
_{k, \,l + 1}}}{{g_{k, \,l + 1}}\,{\bar{g}_{k, \,l}}}}  -
{ \frac {{\bar{f}_{k, \,l}}\,{f_{k, \,l + 1}}}{{
\bar{f}_{k, \,l + 1}}\,{f_{k, \,l}}}}  - {
\frac {{\bar{g}_{k, \,l}}\,{g_{k, \,l + 1}}}{{\bar{g}_{k
, \,l + 1}}\,{g_{k, \,l}}}} \Big)},\\
\label{fcsp-03} && {v_{k + 1, \,l}} - {v_{k, \,l}}={ \frac {{\rm i}a}{2}}
\Big({ \frac {{\bar{f}_{k + 1, \,l}}\,{\bar{f}_{k, \,l}}}{{f_{k + 1, \,l}}\,{f_{k, \,l}}}}  -
{ \frac {{\bar{g}_{k + 1, \,l}}\,{\bar{g}_{k
, \,l}}}{{g_{k + 1, \,l}}\,{g_{k, \,l}}}}  - {
\frac {{f_{k + 1, \,l}}\,{f_{k, \,l}}}{{\bar{f}_{k + 1, \,l}}
\,{\bar{f}_{k, \,l}}}}  + { \frac {{g_{k + 1, \,
l}}\,{g_{k, \,l}}}{{\bar{g}_{k + 1, \,l}}\,{\bar{g}_{k,
\,l}}}} \Big),\\
\label{fcsp-04} && {v_{k, \,l + 1}} + {v_{k, \,l}}={ \frac {\rm i}{2b}}
\,({ \frac {{f_{k, \,l}}\,{\bar{f}_{k, \,l + 1}}
}{{f_{k, \,l + 1}}\,{\bar{f}_{k, \,l}}}}  - {
\frac {{g_{k, \,l}}\,{\bar{g}_{k, \,l + 1}}}{{g_{k, \,l + 1}}
\,{\bar{g}_{k, \,l}}}}  - { \frac {{\bar{f}
_{k, \,l}}\,{f_{k, \,l + 1}}}{{\bar{f}_{k, \,l + 1}}\,{f_{k,
\,l}}}}  + { \frac {{\bar{g}_{k, \,l}}\,{g_{k,
\,l + 1}}}{{\bar{g}_{k, \,l + 1}}\,{g_{k, \,l}}}} ), \\
\label{fcsp-05} && {y_{k + 1, \,l}} - {y_{k, \,l}}={ \frac {a}{2}}
\,\Big({ \frac {{\bar{f}_{k + 1, \,l}}\,{\bar{f}
_{k, \,l}}}{{f_{k + 1, \,l}}\,{f_{k, \,l}}}}    + { \frac {{f_{k + 1,
\,l}}\,{f_{k, \,l}}}{{\bar{f}_{k + 1, \,l}}\,{\bar{f}_{k
, \,l}}}}   \Big),\\
\label{fcsp-06} && {z_{k + 1, \,l}} - {z_{k, \,l}}={ \frac {a}{2}}
\,\Big({  \frac {{\bar{g}_{k + 1, \,l}}\,{\bar{g}_{k, \,l}}}{{g_{k
 + 1, \,l}}\,{g_{k, \,l}}}}    + { \frac {{g_{k + 1, \,l}}\,{g_{k, \,l}}
}{{\bar{g}_{k + 1, \,l}}\,{\bar{g}_{k, \,l}}}} \Big),\\
\label{fcsp-07} && {y_{k, \,l + 1}} - {y_{k, \,l}}= - { \frac {1}{b}}
 + { \frac {1}{2b}} \,{ \Big(
{ \frac {{f_{k, \,l}}\,{\bar{f}_{k, \,l + 1}}}{{
f_{k, \,l + 1}}\,{\bar{f}_{k, \,l}}}}   + { \frac {{\bar{f}
_{k, \,l}}\,{f_{k, \,l + 1}}}{{\bar{f}_{k, \,l + 1}}\,{f_{k,
\,l}}}}  \Big) },\\
\label{fcsp-08} && {z_{k, \,l + 1}} - {z_{k, \,l}}= - { \frac {1}{b}}
 + { \frac {1}{2b}} \,{ \Big(
 {\frac {{g_{k, \,l}}\,{\bar{g}_{k, \,l + 1}}}{{g_{k, \,l + 1}}
\,{\bar{g}_{k, \,l}}}}    + { \frac {{\bar{g}_{k, \,l}}\,{g_{k,
\,l + 1}}}{{\bar{g}_{k, \,l + 1}}\,{g_{k, \,l}}}}\Big) }.
\end{eqnarray}

Then, the ratios on the r.h.s of Eqs. (\ref{fcsp-01})-(\ref{fcsp-08}) can be solved as
\begin{eqnarray}
\label{fcsp-09} &&\frac {{\bar{f}_{k + 1, \,l}}\,{\bar{f}_{k, \,l}}}{{f_{k + 1, \,l}}\,{f_{k, \,l}}}=\frac{1}{a}\Big[y_{k+1,l}-y_{k,l}-\frac{{\rm i}}{2}(u_{k+1,l}-u_{k,l}+v_{k+1,l}-v_{k,l})  \Big],\\
\label{fcsp-10}  &&\frac {{f_{k + 1,\,l}}\,{f_{k, \,l}}}{{\bar{f}_{k + 1, \,l}}\,{\bar{f}_{k, \,l}}}=\frac{1}{a}\Big[y_{k+1,l}-y_{k,l}+\frac{{\rm i}}{2}(u_{k+1,l}-u_{k,l}+v_{k+1,l}-v_{k,l})  \Big],\\
\label{fcsp-11} &&\frac {{f_{k, \,l}}\,{\bar{f}_{k, \,l + 1}}}{{f_{k, \,l + 1}}\,{\bar{f}_{k, \,l}}}=1+b\Big[ y_{k,l+1}-y_{k,l}-\frac{{\rm i}}{2}(u_{k,l+1}+u_{k,l}+v_{k,l+1}+v_{k,l}) \Big],\\
\label{fcsp-12} &&\frac {{\bar{f}_{k, \,l}}\,{f_{k, \,l + 1}}}{{\bar{f}_{k, \,l + 1}}\,{f_{k,\,l}}}=1+b\Big[ y_{k,l+1}-y_{k,l}+\frac{{\rm i}}{2}(u_{k,l+1}+u_{k,l}+v_{k,l+1}+v_{k,l}) \Big],
\end{eqnarray}
\begin{eqnarray}
\label{fcsp-13} &&\frac {{\bar{g}_{k + 1, \,l}}\,{\bar{g}_{k, \,l}}}{{g_{k + 1, \,l}}\,{g_{k, \,l}}}=\frac{1}{a}\Big[z_{k+1,l}-z_{k,l}-\frac{{\rm i}}{2}(u_{k+1,l}-u_{k,l}-v_{k+1,l}+v_{k,l})  \Big],\\
\label{fcsp-14} &&\frac {{g_{k + 1,\,l}}\,{g_{k, \,l}}}{{\bar{g}_{k + 1, \,l}}\,{\bar{g}_{k, \,l}}}=\frac{1}{a}\Big[z_{k+1,l}-z_{k,l}+\frac{{\rm i}}{2}(u_{k+1,l}-u_{k,l}-v_{k+1,l}+v_{k,l})  \Big],\\
\label{fcsp-15} &&\frac {{g_{k, \,l}}\,{\bar{g}_{k, \,l + 1}}}{{g_{k, \,l + 1}}\,{\bar{g}_{k, \,l}}}=1+b\Big[ z_{k,l+1}-z_{k,l}-\frac{{\rm i}}{2}(u_{k,l+1}+u_{k,l}-v_{k,l+1}-v_{k,l}) \Big],\\
\label{fcsp-16} &&\frac {{\bar{g}_{k, \,l}}\,{g_{k, \,l + 1}}}{{\bar{g}_{k, \,l + 1}}\,{g_{k,\,l}}}=1+b\Big[ z_{k,l+1}-z_{k,l}+\frac{{\rm i}}{2}(u_{k,l+1}+u_{k,l}-v_{k,l+1}-v_{k,l}) \Big].
\end{eqnarray}
By making a shift of $l \to l+1$ in (\ref{fcsp-09}), then dividing it by (\ref{fcsp-10}), meanwhile, dividing (\ref{fcsp-11}) by (\ref{fcsp-12}) after a shift of $k \to k+1$,  one obtains
\begin{eqnarray}
\label{fcsp-17} &&\nonumber \frac{y_{k+1,l+1}-y_{k,l+1}-\frac{{\rm i}}{2}(u_{k+1,l+1}-u_{k,l+1}+v_{k+1,l+1}-v_{k,l+1})}{y_{k+1,l}-y_{k,l}+\frac{{\rm i}}{2}(u_{k+1,l}-u_{k,l}+v_{k+1,l}-v_{k,l})}\\
&=& \frac{1+b\Big[ y_{k+1,l+1}-y_{k+1,l}-\frac{{\rm i}}{2}(u_{k+1,l+1}+u_{k+1,l}+v_{k+1,l+1}+v_{k+1,l}) \Big]}{1+b\Big[ y_{k,l+1}-y_{k,l}+\frac{{\rm i}}{2}(u_{k,l+1}+u_{k,l}+v_{k,l+1}+v_{k,l}) \Big]}.
\end{eqnarray}
Similarly, one can obtain
\begin{eqnarray}
\label{fcsp-18} && \nonumber \frac{z_{k+1,l+1}-z_{k,l+1}-\frac{{\rm i}}{2}(u_{k+1,l+1}-u_{k,l}-v_{k+1,l+1}+v_{k,l+1})}{z_{k+1,l}-z_{k,l}+\frac{{\rm i}}{2}(u_{k+1,l}-u_{k,l}-v_{k+1,l}+v_{k,l})}\\
& =& \frac{1+b\Big[z_{k+1,l+1}-z_{k+1,l}-\frac{{\rm i}}{2}(u_{k+1,l+1}+u_{k+1,l}-v_{k+1,l+1}-v_{k+1,l}) \Big]}{1+b\Big[ z_{k,l+1}-z_{k,l}+\frac{{\rm i}}{2}(u_{k,l+1}+u_{k,l}-v_{k,l+1}-v_{k,l}) \Big]}\,,
\end{eqnarray}
from relations (\ref{fcsp-13})-(\ref{fcsp-15}). Equating the real parts and imaginary parts of (\ref{fcsp-17}) and (\ref{fcsp-18}), we have the fully discrete CSP equations (\ref{fcsp-19})--(\ref{fcsp-22}).
\end{proof}

\end{document}